\def\sec{\uppercase\expandafter{\romannumeral\the\sectnum}}
\def\liadot{\hskip-4pt .}
\def\cmp #1{{Commun.  Math.  Phys.} {\bf #1}}
\def\pnas #1{{Proc. Natl. Acad. Sci.} {\bf #1}}
\def\jmp #1{{J. Math. Phys.} {\bf #1}}
\def\pl #1{{Phys. Lett.} {\bf #1}}
\def\np #1{{Nucl. Phys.} {\bf #1}}

\def\ijmp #1{{Int. J. Mod. Phys.} {\bf #1}}

\catcode`\@=\active
\magnification=1200
\def\makeinnocent#1{\catcode`#1=12 }
\tolerance10000

\def\comment{\begingroup
    \let\do\makeinnocent \dospecials
    \makeinnocent\^^L 
    \endlinechar`\^^M \catcode`\^^M=12 \xcomment}
{\catcode`\^^M=12 \endlinechar=-1 %
 \gdef\xcomment#1^^M{\def\test{#1}
      \ifx\test\plainendcommenttest
          \let\next\endgroup
      \else\ifx\test\lalaendcommenttest
          \def\next{\endgroup\end{comment}}
      \else
          \let\next\xcomment
      \fi \fi \next}
}

{\escapechar=-1
 \xdef\plainendcommenttest{\string\\endcomment}
 \xdef\lalaendcommenttest{\string\\end\string\{comment\string\}}
}

\font\titfnt=cmbx10 at 14.40 truept
\newcount\sectnum
\global\sectnum=0
 \newcount\thmnum
\global\thmnum=0
\def\blank{\vskip 12pt}
\def\blankii{\blank\blank}
\def\blankm{\vskip 6pt}
\def\lora{\longrightarrow}

\def\squaro{\mathchoice\sqr35\sqr35\sqr{2.1}5\sqr{1.5}5}
\def\rt#1{\hfill{#1}}
\def\endproof{\rt{  $\squaro$} \par\blankii}

\def\sezione#1{\edef\sectname{#1}\global\advance\sectnum by 1\thmnum=1
\goodbreak\vskip 48pt plus 60 pt\noindent
               {\bf\sec .$\quad$ {#1}}
\immediate\write\fileind
           {\par\hangindent\parindent }\ignorespaces
              \immediate\write\fileind{\line{{\bf\sec.}
$\quad${#1}}\leaderfill\the\count0}\ignorespaces
               \par\blankii\nobreak    \cleareqnum}

\def\prop#1#2{\global\advance\thmnum by 1
        \xdef#1{Proposition \sectnum\the\thmnum}
        \bigbreak\noindent{\bf Proposition \sec.\the\thmnum.}
        {\it#2} }
\newcount\convnum
\convnum=1
\def\convention#1#2{
        \xdef#1{Convention  {\bf\romannumeral\convnum}}
        \bigbreak\noindent{\bf Convention \romannumeral\convnum.}
        {\it#2}\global\advance\convnum by 1 }
\def\definizione#1#2{
        \xdef#1{Definition  \sec.\the\thmnum}
        \bigbreak\noindent{\bf Definition \sec.\the\thmnum.}
        {\it#2}\global\advance\thmnum by 1 }
\def\remark#1#2{
        \xdef#1{Remark  \sec.\the\thmnum}
        \bigbreak\noindent{\bf Remark  \sec.\the\thmnum.}
        {\it#2}\global\advance\thmnum by 1 }

\def\notazione#1#2{
        \xdef#1{Notation  \sec.\the\thmnum}
        \bigbreak\noindent{\bf Notation  \sec.\the\thmnum.}
        {\it#2}\global\advance\thmnum by 1 }

\def\lemma#1#2{
        \xdef#1{Lemma  \sec.\the\thmnum}
        \bigbreak\noindent{\bf Lemma  \sec.\the\thmnum.}
        {\it#2}\global\advance\thmnum by 1}
\def\theorem#1#2{
        \xdef#1{Theorem  \sec.\the\thmnum }
        \bigbreak\noindent{\bf Theorem  \sec.\the\thmnum.}
        {\it#2}\global\advance\thmnum by 1 }
\def\cor#1#2{\global\advance\thmnum by 1
        \xdef#1{Corollary \sec.\the\thmnum}
        \bigbreak\noindent{\bf Corollary \sec.\the\thmnum.}
        {\it#2} }
\def\proof#1{\vskip10pt{\it Proof of
#1}}

\newcount\eqnum
\global\eqnum=0
\def\cleareqnum{\eqnum=0}
\def\num{\global\advance\eqnum by 1
        \eqno({\rm \sec.\the\eqnum)}}

\def\eqalignnum{\global\advance\eqnum by 1
        ({\rm\sec}.\the\eqnum)}

\def\ref#1{\num  \xdef#1{(\sec.\the\eqnum)}}
\def\eqalignref#1{\eqalignnum  \xdef#1{( \sec.\the\eqnum)}}

\def\title#1{\centerline{\bf\titfnt#1}}

\def\today{\ifcase\month\or January\or February\or March\or
        April\or May\or June\or July\or August\or September\or
        October\or November\or December\fi\space\number\day,
        \number\year}

\newwrite\fileref
\immediate\openout\fileref=ref.tmp1
\immediate\write\fileref{\parindent 30pt}
\newwrite\fileack
\immediate\openout\fileack=ref.tmp2
\immediate\write\fileack{\parindent 30pt}

\newcount\norefe

\def\refe{\advance\norefe by 1}
\def\quot#1#2{\refe\xdef#1{[\the\norefe]}[\the\norefe]\immediate\write\fileref
 {\blank\hangindent\parindent}
\ignorespaces\immediate\write\fileref{[\the\norefe]{#2}}\ignorespaces}
\def\ack#1{ \hskip-11pt
\immediate\write\fileack
           {\par\hangindent\parindent}
\ignorespaces
              \immediate\write\fileack{ {#1}}
\ignorespaces}
\def\square {\mathchoice\sqr34\sqr34\sqr{2.1}3\sqr{1.5}3}

\def\Ascr{{\cal A}}
\def\Bscr{{\cal B}}

\def\Dscr{{\cal D}}

\def\Fscr{{\cal F}}
\def\Gscr{{\cal G}}

\def\Pscr{{\cal P}}

\def\Vscr{{\cal V}}
\def\Wscr{{\cal W}}

\def\H{{\hbox{Hol}\;}}
\def\tr{{\hbox{Tr}\;}}
\def\sqr#1#2{{\vcenter {\hrule height.#2pt
       \hbox {\vrule width.#2pt height#1pt \kern#1pt
       \vrule width.#2pt}
       \hrule height.#2pt}}}
\def\leaderfill{\leaders\hbox to 1em{\hss.\hss}\hfill}

\newwrite\fileind
\immediate\openout\fileind=ref.tmp
\immediate\write\fileind{\parindent 30pt}


\def\corollario#1{\goodbreak\vskip 7pt plus 10 pt\noindent
               {{\bf  Corollary   \sec.\the\thmnum}
\quad {\it #1}}
    \global\advance\thmnum by 1}


\newcount\nota

\def\cita{\global\advance\nota by 1}
\def\cito{{\the\nota}}
\def\foot#1{\cita\footnote{$^\cito$}{#1}}

\def\mapdown#1{\Big\downarrow\rlap{$\vcenter{\hbox{$\scriptstyle#1$}}$}}

\def\mapdownn#1
{\bigg\downarrow\rlap{$\vcenter{\hbox{$\scriptstyle#1$}}$}}

\def\riga#1#2{\mathop{ \buildrel #1
 \over{\hbox to #2pt{\rightarrowfill}}}}
\def\liga#1#2{\mathop{ \buildrel #1
 \over{\hbox to #2pt{\leftarrowfill}}}}

%

\catcode`\@=11

\font\tenmsx=msam10
\font\sevenmsx=msam7
\font\fivemsx=msam5
\font\tenmsy=msbm10
\font\sevenmsy=msbm7
\font\fivemsy=msbm5
\newfam\msxfam
\newfam\msyfam
\textfont\msxfam=\tenmsx  \scriptfont\msxfam=\sevenmsx
  \scriptscriptfont\msxfam=\fivemsx
\textfont\msyfam=\tenmsy  \scriptfont\msyfam=\sevenmsy
  \scriptscriptfont\msyfam=\fivemsy

\def\hexnumber@#1{\ifcase#1 0\or1\or2\or3\or4\or5\or6\or7\or8\or9\or
        A\or B\or C\or D\or E\or F\fi }

 \expandafter\ifx\csname frak\endcsname\relax
   \font\teneuf=eufm10
  \font\seveneuf=eufm7
  \font\fiveeuf=eufm5
  \newfam\euffam
  \textfont\euffam=\teneuf
   \scriptfont\euffam=\seveneuf
  \scriptscriptfont\euffam=\fiveeuf
   \def\frak{\ifmmode\let\next\frak@\else
    \def\next{\errmessage{Use \string\frak\space only in math mode}}
\fi
\next
}
   \def\goth{\ifmmode\let\next\frak@\else
  \def\next{\errmessage{Use \string\goth\space only in math mode}}\fi\next}
   \def\frak@#1{{\frak@@{#1}}}
   \def\frak@@#1{\fam\euffam#1}
 \fi
\edef\msx@{\hexnumber@\msxfam}
\edef\msy@{\hexnumber@\msyfam}

\mathchardef\boxdot="2\msx@00
\mathchardef\boxplus="2\msx@01
\mathchardef\boxtimes="2\msx@02
\mathchardef\square="0\msx@03
\mathchardef\blacksquare="0\msx@04
\mathchardef\centerdot="2\msx@05
\mathchardef\lozenge="0\msx@06
\mathchardef\blacklozenge="0\msx@07
\mathchardef\circlearrowright="3\msx@08
\mathchardef\circlearrowleft="3\msx@09
\mathchardef\rightleftharpoons="3\msx@0A
\mathchardef\leftrightharpoons="3\msx@0B
\mathchardef\boxminus="2\msx@0C
\mathchardef\Vdash="3\msx@0D
\mathchardef\Vvdash="3\msx@0E
\mathchardef\vDash="3\msx@0F
\mathchardef\twoheadrightarrow="3\msx@10
\mathchardef\twoheadleftarrow="3\msx@11
\mathchardef\leftleftarrows="3\msx@12
\mathchardef\rightrightarrows="3\msx@13
\mathchardef\upuparrows="3\msx@14
\mathchardef\downdownarrows="3\msx@15
\mathchardef\upharpoonright="3\msx@16

\mathchardef\downharpoonright="3\msx@17
\mathchardef\upharpoonleft="3\msx@18
\mathchardef\downharpoonleft="3\msx@19
\mathchardef\rightarrowtail="3\msx@1A
\mathchardef\leftarrowtail="3\msx@1B
\mathchardef\leftrightarrows="3\msx@1C
\mathchardef\rightleftarrows="3\msx@1D
\mathchardef\Lsh="3\msx@1E
\mathchardef\Rsh="3\msx@1F
\mathchardef\rightsquigarrow="3\msx@20
\mathchardef\leftrightsquigarrow="3\msx@21
\mathchardef\looparrowleft="3\msx@22
\mathchardef\looparrowright="3\msx@23
\mathchardef\circeq="3\msx@24
\mathchardef\succsim="3\msx@25
\mathchardef\gtrsim="3\msx@26
\mathchardef\gtrapprox="3\msx@27
\mathchardef\multimap="3\msx@28
\mathchardef\therefore="3\msx@29
\mathchardef\because="3\msx@2A
\mathchardef\doteqdot="3\msx@2B

\mathchardef\triangleq="3\msx@2C
\mathchardef\precsim="3\msx@2D
\mathchardef\lesssim="3\msx@2E
\mathchardef\lessapprox="3\msx@2F
\mathchardef\eqslantless="3\msx@30
\mathchardef\eqslantgtr="3\msx@31
\mathchardef\curlyeqprec="3\msx@32
\mathchardef\curlyeqsucc="3\msx@33
\mathchardef\preccurlyeq="3\msx@34
\mathchardef\leqq="3\msx@35
\mathchardef\leqslant="3\msx@36
\mathchardef\lessgtr="3\msx@37
\mathchardef\backprime="0\msx@38
\mathchardef\risingdotseq="3\msx@3A
\mathchardef\fallingdotseq="3\msx@3B
\mathchardef\succcurlyeq="3\msx@3C
\mathchardef\geqq="3\msx@3D
\mathchardef\geqslant="3\msx@3E
\mathchardef\gtrless="3\msx@3F
\mathchardef\sqsubset="3\msx@40
\mathchardef\sqsupset="3\msx@41
\mathchardef\vartriangleright="3\msx@42
\mathchardef\vartriangleleft="3\msx@43
\mathchardef\trianglerighteq="3\msx@44
\mathchardef\trianglelefteq="3\msx@45
\mathchardef\bigstar="0\msx@46
\mathchardef\between="3\msx@47
\mathchardef\blacktriangledown="0\msx@48
\mathchardef\blacktriangleright="3\msx@49
\mathchardef\blacktriangleleft="3\msx@4A
\mathchardef\vartriangle="0\msx@4D
\mathchardef\blacktriangle="0\msx@4E
\mathchardef\triangledown="0\msx@4F
\mathchardef\eqcirc="3\msx@50
\mathchardef\lesseqgtr="3\msx@51
\mathchardef\gtreqless="3\msx@52
\mathchardef\lesseqqgtr="3\msx@53
\mathchardef\gtreqqless="3\msx@54
\mathchardef\Rrightarrow="3\msx@56
\mathchardef\Lleftarrow="3\msx@57
\mathchardef\veebar="2\msx@59
\mathchardef\barwedge="2\msx@5A
\mathchardef\doublebarwedge="2\msx@5B
\mathchardef\angle="0\msx@5C
\mathchardef\measuredangle="0\msx@5D
\mathchardef\sphericalangle="0\msx@5E
\mathchardef\varpropto="3\msx@5F
\mathchardef\smallsmile="3\msx@60
\mathchardef\smallfrown="3\msx@61
\mathchardef\Subset="3\msx@62
\mathchardef\Supset="3\msx@63
\mathchardef\Cup="2\msx@64

\mathchardef\Cap="2\msx@65

\mathchardef\curlywedge="2\msx@66
\mathchardef\curlyvee="2\msx@67
\mathchardef\leftthreetimes="2\msx@68
\mathchardef\rightthreetimes="2\msx@69
\mathchardef\subseteqq="3\msx@6A
\mathchardef\supseteqq="3\msx@6B
\mathchardef\bumpeq="3\msx@6C
\mathchardef\Bumpeq="3\msx@6D
\mathchardef\lll="3\msx@6E

\mathchardef\ggg="3\msx@6F

\mathchardef\circledS="0\msx@73
\mathchardef\pitchfork="3\msx@74
\mathchardef\dotplus="2\msx@75
\mathchardef\backsim="3\msx@76
\mathchardef\backsimeq="3\msx@77
\mathchardef\complement="0\msx@7B
\mathchardef\intercal="2\msx@7C
\mathchardef\circledcirc="2\msx@7D
\mathchardef\circledast="2\msx@7E
\mathchardef\circleddash="2\msx@7F
\def\ulcorner{\delimiter"4\msx@70\msx@70 }
\def\urcorner{\delimiter"5\msx@71\msx@71 }
\def\llcorner{\delimiter"4\msx@78\msx@78 }
\def\lrcorner{\delimiter"5\msx@79\msx@79 }
\def\yen{\mathhexbox\msx@55 }
\def\checkmark{\mathhexbox\msx@58 }
\def\circledR{\mathhexbox\msx@72 }
\def\maltese{\mathhexbox\msx@7A }
\mathchardef\lvertneqq="3\msy@00
\mathchardef\gvertneqq="3\msy@01
\mathchardef\nleq="3\msy@02
\mathchardef\ngeq="3\msy@03
\mathchardef\nless="3\msy@04
\mathchardef\ngtr="3\msy@05
\mathchardef\nprec="3\msy@06
\mathchardef\nsucc="3\msy@07
\mathchardef\lneqq="3\msy@08
\mathchardef\gneqq="3\msy@09
\mathchardef\nleqslant="3\msy@0A
\mathchardef\ngeqslant="3\msy@0B
\mathchardef\lneq="3\msy@0C
\mathchardef\gneq="3\msy@0D
\mathchardef\npreceq="3\msy@0E
\mathchardef\nsucceq="3\msy@0F
\mathchardef\precnsim="3\msy@10
\mathchardef\succnsim="3\msy@11
\mathchardef\lnsim="3\msy@12
\mathchardef\gnsim="3\msy@13
\mathchardef\nleqq="3\msy@14
\mathchardef\ngeqq="3\msy@15
\mathchardef\precneqq="3\msy@16
\mathchardef\succneqq="3\msy@17
\mathchardef\precnapprox="3\msy@18
\mathchardef\succnapprox="3\msy@19
\mathchardef\lnapprox="3\msy@1A
\mathchardef\gnapprox="3\msy@1B
\mathchardef\nsim="3\msy@1C
\mathchardef\ncong="3\msy@1D

\mathchardef\varsubsetneq="3\msy@20
\mathchardef\varsupsetneq="3\msy@21
\mathchardef\nsubseteqq="3\msy@22
\mathchardef\nsupseteqq="3\msy@23
\mathchardef\subsetneqq="3\msy@24
\mathchardef\supsetneqq="3\msy@25
\mathchardef\varsubsetneqq="3\msy@26
\mathchardef\varsupsetneqq="3\msy@27
\mathchardef\subsetneq="3\msy@28
\mathchardef\supsetneq="3\msy@29
\mathchardef\nsubseteq="3\msy@2A
\mathchardef\nsupseteq="3\msy@2B
\mathchardef\nparallel="3\msy@2C
\mathchardef\nmid="3\msy@2D
\mathchardef\nshortmid="3\msy@2E
\mathchardef\nshortparallel="3\msy@2F
\mathchardef\nvdash="3\msy@30
\mathchardef\nVdash="3\msy@31
\mathchardef\nvDash="3\msy@32
\mathchardef\nVDash="3\msy@33
\mathchardef\ntrianglerighteq="3\msy@34
\mathchardef\ntrianglelefteq="3\msy@35
\mathchardef\ntriangleleft="3\msy@36
\mathchardef\ntriangleright="3\msy@37
\mathchardef\nleftarrow="3\msy@38
\mathchardef\nrightarrow="3\msy@39
\mathchardef\nLeftarrow="3\msy@3A
\mathchardef\nRightarrow="3\msy@3B
\mathchardef\nLeftrightarrow="3\msy@3C
\mathchardef\nleftrightarrow="3\msy@3D
\mathchardef\divideontimes="2\msy@3E
\mathchardef\varnothing="0\msy@3F
\mathchardef\nexists="0\msy@40
\mathchardef\mho="0\msy@66
\mathchardef\eth="0\msy@67
\mathchardef\eqsim="3\msy@68
\mathchardef\beth="0\msy@69
\mathchardef\gimel="0\msy@6A
\mathchardef\daleth="0\msy@6B
\mathchardef\lessdot="3\msy@6C
\mathchardef\gtrdot="3\msy@6D
\mathchardef\ltimes="2\msy@6E
\mathchardef\rtimes="2\msy@6F
\mathchardef\shortmid="3\msy@70
\mathchardef\shortparallel="3\msy@71
\mathchardef\smallsetminus="2\msy@72
\mathchardef\thicksim="3\msy@73
\mathchardef\thickapprox="3\msy@74
\mathchardef\approxeq="3\msy@75
\mathchardef\succapprox="3\msy@76
\mathchardef\precapprox="3\msy@77
\mathchardef\curvearrowleft="3\msy@78
\mathchardef\curvearrowright="3\msy@79
\mathchardef\digamma="0\msy@7A
\mathchardef\varkappa="0\msy@7B
\mathchardef\hslash="0\msy@7D
\mathchardef\hbar="0\msy@7E
\mathchardef\backepsilon="3\msy@7F

\def\Bbb{\ifmmode\let\next\Bbb@\else
 \def\next{\errmessage{Use \string\Bbb\space only in math mode}}\fi\next}
\def\Bbb@#1{{\Bbb@@{#1}}}
\def\Bbb@@#1{\fam\msyfam#1}

\catcode`\@=\active

\def\incul{\hookrightarrow}
\def\rsa{\rightsquigarrow}
\hfuzz500pt
\def\dota{ {\buildrel \circ\over{ \!  A}}}

 \def\dss{\mathop{  \left.{d\over d s}\right\vert _{s=0 }  }}

\def\st{\mathop{\hbox{such that}\;\;}}
\def\dy{\displaystyle}
\def\ad{\mathop{{\rm{ad}}}}
\def\Ad{\mathop{{\rm{Ad}}}}

\def\Adgm{\mathop{
{\rm{Ad}}_{g^{-1} }
}}
\def\adP{\mathop{{\rm{ad}} \, P}}
\def\tadP{\mathop{{\rm{ad}} \, P\oplus {\rm{ad}} \, P}}
\def\adTP{\mathop{{\rm{ad}} \, \T P}}

\def\tharm#1{\mathop{\widetilde{\hbox{Harm}}_A^{#1}(M,\adP)}}
\def\harm#1{\mathop{\hbox{Harm}_A^{#1}(M,\adP)}}
\def\form#1{\mathop{\Omega^{#1}(M,\adP)}}
\def\Tform#1{\mathop{\Omega^{#1}(\T M,\adTP)}}
\def\tform#1{\mathop{\Omega^{#1}(M,\adP)\oplus\Omega^{#1}(M,\adP) }}
\def\ttform#1{\mathop{\Omega^{#1}(M,\tadP)}}

\def\hexnumber@#1{\ifcase#1
0\or1\or2\or3\or4\or5\or6\or7\or8\or9\or
        A\or B\or C\or D\or E\or F\fi }
\font\tencmmib=cmmib10 \font\sevencmmib=cmmib10 scaled \magstep0
\font\fivecmmib=cmmib5
 \skewchar\tencmmib'177 \skewchar\sevencmmib'177 \skewchar\fivecmmib'177
 \newfam\cmmibfam
\textfont\cmmibfam\tencmmib
 \scriptfont\cmmibfam\sevencmmib
\scriptscriptfont\cmmibfam\fivecmmib
 \font\tencmbsy=cmbsy10 \font\sevencmbsy=cmbsy7 \font\fivecmbsy=cmbsy5
 \skewchar\tencmbsy'60 \skewchar\sevencmbsy'60 \skewchar\fivecmbsy'60
  \newfam\cmbsyfam
\textfont\cmbsyfam\tencmbsy
 \scriptfont\cmbsyfam\sevencmbsy \scriptscriptfont\cmbsyfam\fivecmbsy

\edef\cmbsy@{\hexnumber@\cmbsyfam}
 \edef\cmmib@{\hexnumber@\cmmibfam}

\mathchardef\balpha="3\cmmib@0B
\mathchardef\bbeta="3\cmmib@0C
\mathchardef\bgamma="3\cmmib@0D
\mathchardef\bdelta="3\cmmib@0E
\mathchardef\bepsilon="3\cmmib@0F
\mathchardef\bzeta="3\cmmib@10
\mathchardef\beeta="3\cmmib@11
\mathchardef\btheta="3\cmmib@12
\mathchardef\biota="3\cmmib@13
\mathchardef\bkappa="3\cmmib@14
\mathchardef\blambda="3\cmmib@15
\mathchardef\bmu="3\cmmib@16
\mathchardef\bnu="3\cmmib@17
\mathchardef\bxi="3\cmmib@18
\mathchardef\bpi="3\cmmib@19
\mathchardef\brho="3\cmmib@1A
\mathchardef\bsigma="3\cmmib@1B
\mathchardef\btau="3\cmmib@1C
\mathchardef\bupsilon="3\cmmib@1D
\mathchardef\bphi="3\cmmib@1E
\mathchardef\bchi="3\cmmib@1F
\mathchardef\bpsi="3\cmmib@20
\mathchardef\bomicron="3\cmmib@21
\mathchardef\bepsilon="3\cmmib@22
\mathchardef\bvartheta="3\cmmib@23
\mathchardef\bvarzeta="3\cmmib@26
\mathchardef\bvarphi="3\cmmib@27
\mathchardef\bbo="3\cmmib@
\def\T{{\rm {T}}}
\def\V{{\rm {V}}}
\def\H{{\rm {H}}}
\def\Id{{\Bbb {Id}}\;}
\def\Gaff{{\Gscr_{{\rm aff}}}}
{\baselineskip=12pt
\nopagenumbers
\line{\hfill \bf HUTMP 97/B361}
\line{\hfill \bf \today}
\vfill
\title{BRST symmetries}
\medskip
\title{for the tangent gauge group}
\vskip1in
\centerline{ {\bf Alberto S. Cattaneo},$^{(a)}$\footnote{$^1$}
{Supported by I.N.F.N. grant No.\ 5565/95
and DOE grant No.\ DE-FG02-94ER25228, Amendment No.\ A003.
}}
\centerline{{\bf Paolo Cotta-Ramusino},$^{(b)}$\footnote{$^2$}
{Supported in part by research grants of the
Ministry of University and Scientific Research (MURST)}
and {\bf Maurizio Rinaldi}$^{(c)}$ {$^2$}}
\vskip12pt
\centerline{$^{(a)}$Lyman Laboratory of Physics}
\centerline{Harvard University}
\centerline{Cambridge, MA 02138, USA}
\vskip12pt
\centerline{$^{(b)}$Department of Mathematics}
\centerline{University of Milano}
\centerline{Via Saldini 50 }\centerline{20133 Milano, Italy}
\centerline{and}
\centerline{I.N.F.N., Sezione di Milano}
\vskip12pt
\centerline{$^{(c)}$Department of Mathematics}
\centerline{University of Trieste}
\centerline{Piazzale Europa 1}
\centerline{34127 Trieste, Italy}
\vfill
\noindent
{\bf Abstract.}
For any principal bundle $P$, one can consider the
subspace of the space of connections
on its tangent bundle $\T P$
given by the tangent bundle $\T\Ascr$
of the space of connections $\Ascr$
on $P$. The tangent
gauge group acts freely on $\T\Ascr$.
Appropriate BRST operators
are introduced for quantum field theories
that include as fields elements of $\T\Ascr$, as well as
tangent vectors to the space of curvatures.
As the simplest application, the BRST symmetry
of the so-called $BF$-Yang--Mills
theory is described and the relevant gauge fixing conditions
are analyzed. A brief account on the topological $BF$ theories is
also included and the relevant Batalin--Vilkovisky operator is
described.
\blank\blank
}
\vfill\eject
\sezione{Introduction}
In this paper we study  the BRST complexes arising from  the action
of
the (iterated) tangent
bundle of the group of gauge transformations $\Gscr$ on the (iterated)
tangent bundle of the space of connections $\Ascr$.

Our main original motivation for such a study was
the understanding of the BRST and the Batalin--Vilkovisky
(BV) structure of
4-dimensional $BF$ theories. These topological theories
have been  shown \quot\bfseven{A. S. Cattaneo, P. Cotta-Ramusino, F. Fucito,
M. Martellini, M. Rinaldi, A. Tanzini, M. Zeni: {  ``Four-Dimensional
Yang--Mills
Theory as a Deformation of Topological BF Theory"}, preprint HUTMP-97/B362-IFUM
547/FT-ROM2F/96/61-hep-th/9705123;} to
be strictly related  to the 4-dimensional
Yang--Mills theory.

Let us first recall what $BF$ action functionals are
in both three and four dimensions.
We consider ``space--time" to be given
by a closed, oriented  Riemannian manifold $M$
with a gauge structure given by a $G$-principal bundle $P$ with base $M$.
The group $G$ is a simple compact connected
Lie Group (typically $SU(N)$) with Lie algebra denoted by
${\frak g}$. Unless otherwise stated, the forms we
are considering on $M$ will take values in the adjoint bundle and hence
their local expression will take  values in ${\frak g}$.

When $M\equiv M^3$ is a {\it 3-dimensional} space, then the most known
topological
theory is that defined in terms of the Chern--Simons action functional
$$CS(A)\equiv \int_{M^3}
\tr \left(A\wedge dA + {1\over 3}A\wedge [A,A]\right),$$
where $A$ is (the local expression of) a
connection on $P(M,G)$.

We can also introduce a field $\widehat B$ given by
a 1-form. The 1-form
$A+t\widehat B$ is again a connection.

The 3-dimensional topological
$BF$ action functional is then defined in terms of
the Chern--Simons action functional
by
$$S_{BF,t}(A,\widehat B)\equiv {1\over{2t}}\bigg(CS(A+t\widehat B)-
CS(A-t\widehat B)\bigg)
=$$
$$\int_{M^3} \tr\left( 2(\widehat B\wedge F_A)
+ {t^2\over 3} \hat B\wedge [\hat B, \hat B]\right),\ref\bfcc $$
where $F_A$ denotes the curvature of $A$ and $t$ is a real parameter.
The action functional \bfcc\ is called the {\it $BF$ action functional
with cosmological constant} \quot\jmpquot{A. S.
Cattaneo,  P. Cotta-Ramusino and M.
Martellini: ``Three-Di\-men\-sion\-al $BF$ Theories and the
Alexander--Conway Invariant of Knots'',
\np{B 346}, 355--382 (1995);
A. S. Cattaneo, P. Cotta-Ramusino, J. Fr\"ohlich and
M. Martellini:
``Topological $BF$ Theories in 3 and 4 Dimensions",
\jmp{ 36}, 6137--6160 (1995);
A. S. Cattaneo: ``Cabled Wilson Loops in $BF$ Theories",
\jmp {37}, 3684--3703 (1996);
A. S. Cattaneo: ``Abelian $BF$ Theories and Knot Invariants",
preprint HUTMP-95/B355, hep-th/9609205,
to appear in \cmp{};}\liadot

In the limit $t\to 0$ we obtain, up to a factor 2,
the so called {\it pure $BF$
action functional}
$$S_{BF}= \int_{M^3} \tr \left(\hat B\wedge F_A\right).$$
One can define observables related to links in $M^3$ for both
$S_{BF,t}$ and $S_{BF}$; it has been shown in \jmpquot\
that the corresponding
expectation values give the coefficients of the Homfly--Jones
polynomials in the first case and are related to the Alexander--Conway
polynomials in the second case.

In {\it 4 dimensions} we first consider
the topological Chern action
functional $$\int_M \tr \bigg(F_A\wedge F_A\bigg).$$
Then we consider a field $B$ given by
a 2-form.
The 4-dimensional topological {\it $BF$ action functional
(with cosmological constant)}
is then defined as
$$S_{BF,t} \equiv {1\over 2t}\int_M \tr \bigg((F_A +t B)\wedge(F_A +t B)
- F_A\wedge F_A\bigg) =$$
$$\int_M \tr \left( B\wedge F_A + {t\over 2}B\wedge B\right),\ref\bftop$$
and again the {\it pure $BF$ action functional $S_{BF}$} is defined
as the limit of the above
one when $t$ goes to $0$.

In 4-dimensions one may also start with the Yang--Mills action functional
$$S_{YM}=\int_M \tr \bigg(F_A \wedge *F_A\bigg),$$
where ${}*$ is the Hodge operator. (We
omit the coupling constant here.)

The corresponding $BF$-Yang--Mills ($BFYM$) action functional reads
$$S_{BFYM,t}\equiv \int_M \tr\left( B\wedge F_A +{t\over 2} B\wedge
*B\right),\ref\bfym$$ whose limit $t\to 0$ gives again
the pure $BF$ action functional.

When $t=1$,  \bfym\ is {\it equivalent, by Gaussian integration,
to the Yang--Mills action functional}.

When $t=0$,  \bfym\ has an extra-symmetry, namely, the translation $B\lora
B+d_A\tau$ where 
$\tau$ is a 1-form.  In order to extend this symmetry for a generic $t$ and
carry on the perturbative
expansion around the critical solutions for the $BF$-Yang--Mills
theory, it is necessary to
deform the action functional \bfym\ into the following action functional:
$$S_{BFYM,\eta,t}\equiv \int_M \tr\left(B\wedge F_A +{t\over 2} (B-d_A\eta)
\wedge
*(B-d_A\eta)\right).\ref\bfymeta$$
Here $\eta$ is a 1-form and the new symmetry
reads, for a generic 1-form $\tau$
$$B\lora B+  d_A \tau,\quad \eta \lora \eta +\tau.\ref\tautransl$$
It is now possible to show that in perturbation theory, the
action functional \bfymeta\ is equivalent to
the Yang--Mills action functional and we refer
to \bfseven\   for such a discussion.

By now it is clear,
just by inspecting all the action functionals written above,
that in both the 3- and the 4-dimensional $BF$ theories, the field
$\hat B$
and respectively
$B$ belong to some tangent space to the space of fields:
the connections in three dimensions and the curvatures in four.

Some
care is to be applied in four dimensions, for the space of ``curvatures"
is not the space of fields where one performs the functional integration.

Specifically, in the $BFYM$ theories with action functional \bfymeta,
one works with triples $(A,\eta,B)$ where
the pair $(A,\eta)$ belongs to the tangent bundle
of the space of connections and $B$ is a tangent vector at $F_A$
to the vector space of 2-forms.

The relevant BRST cohomology
should accomodate both the gauge invariance
and the invariance under translations \tautransl; the latter condition
implies that
eventually no physical quantity should depend on the field $\eta$; that is,
the theory should behave like a topological field theory as far as the field
$\eta$ is concerned.

The existence of such a BRST cohomology is a priory not obvious; the purpose of
this paper is to describe in details such a BRST cohomology, which is nothing
but a straightforward
extension of the BRST cohomology arising from the action of the tangent
gauge group on the tangent bundle of connections.

The plan of the paper is as follows:

In section {\bf II} we recall some aspects of the topological field theories
of cohomological type as discussed in \quot\bs{L. Baulieau and
I. M. Singer: ``Topological Yang--Mills Symmetry",
\np  {  B }(Proc. Suppl.) {\bf 5B}, 12
(1988);}\liadot

In sections {\bf III} and {\bf IV} we study the space of connections
of the tangent bundle of a principal bundle (tangent connections)
and the relevant gauge group.

In section {\bf V} and {\bf VI} we discuss the BRST cohomology
arising from the action of the tangent gauge group on the tangent
bundle of connections.

In section {\bf VII} we show that it is possible to
extend the above BRST operator to the 2-form $B$.

In section {\bf VIII} we show that
the construction can be indefinitely iterated by including $2^n$
2-forms $B_i$, $2^n$ 1-forms $\eta_i$
and the relevant ghosts.

In section {\bf IX} we consider the BV construction corresponding
to the action functional \bftop.

In section {\bf X} we discuss the
orbit space and the moduli space relevant to the action functional
\bfymeta.

Some of the results
of sections {\bf VI}, {\bf VII}, {\bf X} have been anticipated
in \bfseven.

As a final remark, we notice that the
2-form $B$ in 4-dimensional $BF$ theories
can be interpreted geometrically as a tangent vector to the space
of connections over the following bundle.

Take the space $\Pscr_A(P)$ of (free) horizontal paths on $P$ with respect
to a connection $A$. This is
a principal $G$-bundle, whose base space is the space of
the free paths on $M$. The initial-point map $ev_0:\Pscr_A(P)\to P$
is a bundle morphism, so $ev^*_0A$ is a connection on $\Pscr_A(P)$.
Any form $B\in\form{2}$, once integrated over the paths, represents a
tangent vector to the space of connections on $\Pscr_A(P)$.
The computation of the relevant holonomies allows us
to define observables
corresponding to imbedded tori in a (simply connected) 4-manifold $M$, for
which expectation values with respect to the different $BF$ action functionals
can be computed.
We refer to a separate paper
\quot\paths{A. S. Cattaneo, P. Cotta-Ramusino and M. Rinaldi:
``Connections on Loop and Path Spaces and Four-Dimensional
$BF$ Theories", in preparation;} for such a discussion.
\vfill\eject
\sezione{Symmetries in gauge and
topological field theories}

In this section we review some well known facts concerning gauge invariance
in topological (cohomological) theories
and in Yang--Mills theories and the BRST structure of such theories.

The main symmetry group we are considering is the gauge
group $\Gscr$ that acts on the space $\Ascr$
of (irreducible) smooth connections
on $P$.  We assume that we have divided $\Gscr$
by its center, so that
$\Gscr$ acts freely on $\Ascr$ yielding the principal bundle
of gauge-orbits.

As usual, we denote by $\adP$ the associated bundle
$P\times_{Ad}{\frak g}$ and by $\form{*}$ the graded Lie algebra of tensorial
form on $P$ of the adjoint type, or equivalently, of forms on $M$
with values in $\adP$.
For any
$A\in \Ascr$ the covariant exterior derivative will be a linear map
$$d_A\colon\form{*} \to \form{*+1} .$$

The space $\Ascr$ is an affine space and its tangent space
$\T_A\Ascr$ is
the vector space $\form{1}$.

A quantity (observable, action functional)
is gauge invariant if it is invariant
under the action of the group $\Gscr$.
It is moreover independent of the connection if it is invariant under
the translation group $\form{1}$. In quantum field theories one
may wish to consider simultaneously both kinds of invariance.

In this case one is naturally led to consider
the semidirect product
$$\Gscr_{\T} \triangleq \Gscr\ltimes \form{1}.\ref\gt$$
In this group the product of two elements $(g_1,\rho_1)$ and
$(g_2,\rho_2)$ is defined by
$$(g_1,\rho_1)(g_2,\rho_2)= \left(g_1g_2, {\Ad}_{g_2^{-1}} \rho_1+
\rho_2 \right).$$
Here $g_i\in \Gscr$ and $\rho_i\in\form{1}$.

The group $\Gscr_{\T}$ acts on the space of connections
as follows:
$$A\cdot (g,\rho)=A^g + \rho \ref\transl,$$
where $A^g$ denotes the gauge-transformed connection.

This action is not free since $(g, A-A^g)\in \Gscr_{\T}$ leaves the connection
$A$ fixed. At the infinitesimal level, we see that
the freedom of the action is missing since we have non-trivial
solutions of the equation
$$\rho +d_A\chi=0,\quad \rho\in\form{1}, \chi\in\form{0},$$
where $d_A$ is the covariant exterior derivative.

{}From a field-theoretical point of view,
dealing with a symmetry that does not correspond to a
free action is troublesome since a
gauge fixing mechanism and a consistent definition of
ghost fields may not be available.

The cure for this problem is to require $\rho$ to
belong to a complementary
space of the image of $d_A\colon\form{0}\to \form{1}$. In other
words we have to choose a {\it connection on the bundle of gauge orbits}
$$\Ascr\mapsto {\Ascr\over\Gscr}.\ref\gauorb $$

Choosing such a connection is also
called {\it fixing the gauge.}
We denote the above connection on the bundle of gauge orbits by
the symbol $c$.

In turn any connection $c$ on $\Ascr$, determines a connection on the bundle
$${{P\times \Ascr}\over{\Gscr}}\mapsto M\times {\Ascr\over \Gscr}.\ref\peg$$
In fact the 1-form on $P\times \Ascr$ given by
$$(A+c)_{(p,A)}(X, \eta)\triangleq A(X)_p + c(\eta)_{(A,p)},\quad
X\in \T_pP,\, \eta\in \T_A\Ascr\ref\connpeg$$
defines a $\Gscr$-invariant connection on $P\times \Ascr$
\quot\Atia{M.F. Atiyah, I.M. Singer: ``Dirac Operators coupled to Vector
Potentials", \pnas{81}, 2597--2599 (1984);}, i.e.,
a connection on the principal $G$-bundle \peg\ .

The exterior derivative on forms on $P\times \Ascr$ which are of order
$(k,s)$ (i.e., of order $k$ as forms on $P$ and order $s$ as forms on
$\Ascr$) can be written as
$d_{tot} = d +(-1)^k\delta$ where $d$ and $\delta$ are the exterior
derivatives on $P$ and $\Ascr$ respectively.
Analogously  we set the following convention for
the commutator $$[\omega_1,\omega_2]=(-1)^{k_1k_2 + s_1s_2+1}
[\omega_2,\omega_1],$$
where $(k_i,s_i)$ denotes the order of the form $\omega_i$.
As a consequence we have the following expression for the total
exterior covariant derivative
$$d_{A+c}= d_A + (-1)^k \delta_c,\ref\signcov$$
where $k$ denotes again the order of the form on $P.$
In particular we have to remember equation \signcov\
when computing curvatures and their Bianchi identity.

The
order of a form with respect to the space $\Ascr$ is called the
{\it ghost number}.

The operator $\delta$ (or some of its
restrictions or extensions)
is as an example of what in quantum field theory is
called a BRST (or a BRS) operator;
it describes the symmetry of the fields that one considers in the theory.

We have in mind essentially two cases.

The first case is that of the
{\it topological field theory}
\`a la Baulieau--Singer \bs\ or,
according to Witten's terminology \quot\witt{E.Witten:
``Introduction to Cohomological Field Theories",
\ijmp{A6}, 2775--2792 (1991);
}, of a {\it cohomological
field theory}. In this case the theory has no degrees of freedom
and the action functional is invariant under (infinitesimal)
translations of the space
of fields. The corresponding BRST operator is then the exterior derivative
on the space of fields ($\Ascr$ in the previous case).

The second case is that
of a field theory (generally not topological) in which
there is a symmetry group acting on the space of fields and the theory
is only required to be invariant under such a symmetry group.
In this last case the BRST operator is just the restriction of the exterior
derivative on the space of fields to the orbits of that symmetry group.

In the Yang--Mills theory, the BRST operator is the exterior derivative
along the gauge orbits.

Mixed cases will also (and especially) be
considered  in this paper, namely, theories
for which the BRST operator is the exterior derivative for some of the
fields
and the derivative along the fibers (orbits) for other fields.

We may say that these theories are
``semi-topological" or topological in some fields
and non-topological in others. We
use this terminology, even though topological
field theories are not only those
of cohomological type.

In a topological theory (of the cohomological type) defined for
an even-dimensional space,
the typical action functional
is defined in terms of characteristic classes,
so is independent of the choice of the connection.

Let us come back to the Baulieu--Singer theory. Such  a theory
is invariant under translations in the space of
connections, the action functional is
represented by characteristic classes, and
the field
equations  are obtained by
considering the structure equation for the connection \connpeg\ and
the relevant Bianchi identities.

The structure equations read
$$F_A + \psi +\phi = d_AA + d_Ac + {1\over 2} [c,c] - \delta A +\delta c,\ref
\structure$$
where the l.h.s. gives the various components of the
curvature of the connection \connpeg; $F_A$, $\psi,\phi$ are
respectively forms of degree $(2,0)$, $(1,1)$ and $(0,2)$ on the product space
$$P\times \Ascr.$$

The previous equation  can be rewritten as
$$\delta A= d_A c - \psi,\quad \delta c = -{1\over 2}[c,c] +
\phi.\ref\brstpeg$$

Notice that the form $\psi$
is minus the projection of $\form{1}$ onto the
horizontal subspace at $A.$

The Bianchi identity defines the transformation properties of
$\psi$ and $\phi$ (and $F_A$)

$$\delta \psi= d_A\phi -[\psi,c], \quad \delta\phi=[\phi,c]\ref \brstpegi$$
$$d_AF_A=0, \quad \delta F_A =[F_A,c]-d_A\psi.\ref\brstpegii$$

By comparing the first equation in \brstpeg\ with the action
\transl, we see that we have successfully
turned around the problem of the non-freedom
of the action \transl.

As was recalled before,
the {\it Yang--Mills} theory, as opposed to the previous topological theory,
involves the
restriction of the transformation \brstpeg\ to
the orbit passing through $A\in\Ascr$.
The relevant
equations are then
$$\delta A= d_A c ,\quad \delta c = -{1\over 2} [c,c],\ref\brs $$
where $c$ in \brs\ is now the {\it Maurer--Cartan form} on $\Gscr,$ and $\delta$ is the
exterior derivative on $\Gscr$ or, more precisely, on the $\Gscr$-orbit
passing through $A$.

The equations \brs\ are the classical BRST equations \quot\brs{L.
 Bonora and P. Cotta-Ramusino: ``Some Remarks on BRS
Transformations, Anomalies and the Cohomology of
the Lie Algebra of the Group
of Gauge Transformations", \cmp{87}, 589--603 (1983);
C. Becchi, A. Rouet
 and R. Stora: ``Renormalization of
the Abelian Higgs--Kibble Model", \cmp{42},
127 (1975);
I. V. Tyutin, Lebedev Institute preprint N39, 1975;}.

A final remark is in order. Whenever we have a non-free action
of a group over a manifold, one is naturally led to consider the corresponding
equivariant
cohomology. In our case the group is $\Gscr_{\T}$ with its non-free action on
$\Ascr$.
But $\Gscr_{\T}$ is the semidirect product of
an abelian (contractible) subgroup $\form{1}$ times
$\Gscr$, which behaves in
many ways like
a ``compact group" and
acts freely on (the contractible
space) $\Ascr$.

So one has just to consider this last action, or the cohomology of the
principal bundle \peg\ (which in turn
is just the equivariant cohomology corresponding
to the non-free action of $\Gscr$ on $P$).

In the following sections
we will extend this procedure by considering the action of the
tangent bundle of the gauge group
(which also is a Lie group) on the tangent space to the space of connections.
This may look irrelevant from a purely topological point of
view since, as will be recalled below, the tangent gauge group
is again the semidirect product of $\Gscr$ times
an abelian contractible group. This implies that the classifying
spaces for the gauge group and for its tangent bundle are
homotopically equivalent.

But the BRST structure of the tangent gauge group will show
an interesting
degree of flexibility, due exactly to the fact that there
is a large abelian normal subgroup of the tangent gauge group.
This will allow us to include into the BRST algebra the
2-form fields $B$ of the 4-dimensional $BF$ theories.

However, before doing so, we need to discuss some basic facts of the
differential geometry of the tangent bundle of a principal bundle.
\sezione{Tangent principal bundles and their gauge groups}

Let $G$ be a Lie group and let ${\frak g}$ its Lie algebra.
The tangent bundle
$\T G$ is a Lie group with respect to the multiplication given
by the tangent map of the multiplication $m\colon G\times G \mapsto G$
$$
\T m\colon
\T G\times \T G \mapsto \T G
$$
Explicitly this multiplication is given by
$$
(k,u_k)(h,v_h)\equiv (kh, kv_h + u_kh)\quad k,h\in G; u_k\in
\T_kG; v_h \in \T_hG.
$$
Here the right and left multiplications
of a vector by an element of $G$ denote the push-forward of the corresponding
multiplications in $G$.

Since we can represent any vector $u_k$ as $u_k=kx$ for a unique
$x\in {\frak g}=\T_eG$, there is an isomorphism
$$
\T G{\buildrel{\sim}\over{\to}} G\ltimes {\frak g} , \quad (k,u_k)\rsa(k,k^{-1}u_k)
\ref\iotaeq,$$
and on the semidirect product
$G\ltimes {\frak g}$ the product of two elements is defined as
$$
(k,x)(h,y)\equiv (kh,{\Ad}_{h^{-1}}(x) +y)\quad k,h\in G;\; x,y\in {\frak g}.
$$
The inverse of the element
$(k,x)\in (G\ltimes{\frak g})$ is given by
$$
(k,x)^{-1}=(k^{-1},-{\Ad}_{k }( x)),
\ref\inverse
$$
and the conjugation on $G\ltimes {\frak g}$ is given by
$$
 (k,x)(h,y)(k,x)^{-1}=\left(khk^{-1}, {\Ad}_{k }
{\Ad}_{h^{-1}}(x) + {\Ad}_{k }(y)-{\Ad}_{k }(x)
\right).\ref\conjug
$$
\comment
In order to compute
the 1-parameter subgroup generated by $(x,y)\in {\frak g} \times {\frak g}$,
we consider for any $x,y\in {\frak g}$ the curve $\gamma_{x,y}(t)$
in ${\frak g}$ given by
$$\gamma_{x,y}(t)=\displaystyle{\int_0^t}{\Ad}_{\exp(-sx)}y \,ds.$$
The one parameter subgroup generated by $x,y\in{\frak g}$ is then given by
$$
\exp \left(t(x,y)\right)=\left(\exp(tx),\gamma_{x,y}(t)\right).
\ref\onepar
$$
\endcomment
The adjoint action of $G\ltimes {\frak g}$ on
$\T_{(e,0)}(G\ltimes {\frak g})=\frak
g\times\frak g$ is given by
$$
{\Ad}_{(k,z)}(x,y)=\left({\Ad}_k(x), {\Ad}_k([z,x]+y)\right),
\ref\Adeq$$
whereas the commutator  is given by
$$
[(x_1,y_1),(x_2,y_2)]=\left( [x_1,y_2], [x_1,y_2]+[y_1,x_2]\right).
\ref\commut
$$
So $Lie(\T G)$ is isomorphic as a Lie algebra to the semidirect sum
 ${\frak g}\oplus_s{\frak g}.$ We now consider
a $G$-principal bundle $P\mapsto M$, with base space
$M$.  Its tangent bundle is
a $\T G$-principal bundle with right action of $\T G$ on $\T P$
given by the tangent map to the right action
$R\colon P\times G\mapsto P$,
$$\eqalign{
\T  R\colon \T P\times \T G&\mapsto \T P\cr
\left((p,X),(k,u_k)\right)&\rsa \left(pk, Xk + pu_k\right).\cr}
\ref\tptg
$$
Here and in what follows, $Xk$ and $pu_k$ (with $p\in P$, $X\in \T_pP$,
$k\in G$, $u_k\in \T_kG$) are simplified notations for the
partial tangent maps $\T_1
R\colon\T P\times G \to
\T P$ and $\T_2 R\colon P\times \T G\to \T P$.
   We now rewrite the previous action, by
considering the isomorphism
\iotaeq. If $(k,x)\in G\ltimes {\frak g}$,
and $i(x)$ denotes the fundamental
vector field in $P$ generated by $x\in {\frak g}$, then
the action \tptg\ becomes
$$
(p,X)(k,x)= \left(pk,  Xk + \left. i(x)\right|_{pk}\right).
\ref\princ
$$
It is easy to check directly that \princ\ defines a good action
$$
\left[(p,X)(k,x)\right](h,y)= \left(pkh,(Xk)h +  (\left. i(x)
\right |_{pk})h +\left. i(y)\right|_{pkh}\right)$$ $$=
\left(pkh, X(kh) +\left. i\left({\Ad}_{h^{-1}}x\right)\right|_{pkh}
+\left. i(y)\right|_{pkh}\right) = (p,X)\left[(k,x)(h,y)\right].$$
The action \princ\ is free and
moreover $\pi\colon\T P\mapsto \T M$ is a $(G\ltimes {\frak g})$-principal fiber
bundle.

We now shift to the infinite-dimensional case.
We consider the tangent bundle $\T\Gscr$
of the group of gauge transformations.

This can be identified with the semidirect product
of $\Gscr\ltimes \form{0}$, where the product of two elements
is defined by
$$(g_1,\zeta_1)(g_2,\zeta_2)=
\left(g_1g_2, {\Ad}_{g_2^{-1}}\zeta_1+ \zeta_2\right).\ref\tgg$$
Its Lie algebra is given by the semidirect sum
$\form{0} \oplus_s \form{0}$ whose commutator
is defined as follows
$$ [(\chi_1,\zeta_1),(\chi_2,\zeta_2)]=\left([\chi_1,\chi_2], [\chi_1,\zeta_2]+
[\zeta_1,\chi_2]\right).$$

The tangent map to the action of $\Gscr$ on $\Ascr$ defines the action of
 $\T\Gscr$
on
$\T\Ascr\sim\Ascr\times
\form{1}$ as follows
$$(A,\eta)\cdot (g,\zeta)=(A^g, {\Ad}_{g^{-1}}(\eta) + d_{A^g} \zeta).\ref\taction$$
The fundamental vector field at $(A,\eta)\in \T\Ascr$ corresponding
to $(\chi,\zeta)\in Lie (\T\Gscr)$ is given by
$$\left(d_A\chi, [\eta,\chi]+d_A \zeta\right).\ref\fundvf $$

The action \taction\
is free when we understand that only irreducible
connections on $P$ are taken into account and that the group $\Gscr$ has
been divided by its center.

A {\it gauge transformation} for $\T P$ is, by definition, a map
$$
f\colon\T P\lora G\ltimes {\frak g} \quad \st f[(p,X)(k,x)]=(k,x)^{-1}f(p,X)(k,x).
$$
We have the following
\theorem\tgauge {The tangent bundle $\T\Gscr$ of the gauge Lie group $\Gscr$
is a proper subgroup of
the group $\Gscr_{\T P}$ of gauge transformations for $\T P$.}
\proof\tgauge

We consider the tangent map to the evaluation
map $ev\colon P\times \Gscr \to G$ at $(g,\chi)\in \Gscr\ltimes\form{0}
\sim \T \Gscr.$

It is a map
$$\T P\lora G\ltimes {\frak g}$$ given by
$$\left(g(p),
g^{-1}dg(p,X) + \chi(p) \right)\in G\ltimes{\frak g},$$
where we have to remember that the logarithmic derivative
$g^{-1}dg$ is a map defined on $\T P$ with values in ${\frak g}.$

In order to show that it is a gauge transformation for $\T P$,
it is enough to see
that for any $(p,X)\in \T P$ and $(k,x)\in G\ltimes {\frak g}$
we have the equation
$$g^{-1}dg\left\{(p,X)(k,x)\right\}=
g^{-1}dg\left\{\left(pk,  Xk + i(x)|_{pk}\right)\right\}=$$
$$
{\Ad}_{k^{-1}}[g^{-1}dg (p,X)] +x -{\Ad}_{k^{-1}}
{\Ad}_{g^{-1}(p)} {\Ad}_{k}x=p_2\left({\Ad}_{(k,x)^{-1}}\left\{g^{-1}dg(p,X)
\right\}\right),
$$
where $p_2:{\frak g}\oplus_s{\frak g}\to {\frak g}$ is the projection onto the
second component.
For any $\eta\in\form{1}$
and for any $(g,\chi)\in \Gscr\ltimes Lie(\Gscr)$ the map
$$(p,X)\rsa \left(g(p), g^{-1}dg(p,X) + \chi(p)
+ \eta(p,X) \right)$$
is also a gauge transformation on $\T P$ and this shows that the inclusion
$\T\Gscr\incul \Gscr_{\T P}$ is proper.
\endproof
We have also shown that the group $\Gscr_{\T P}$ includes
the group $\Gaff$ of {\it affine gauge transformations},
defined as the semidirect product
$\T\Gscr \ltimes \form{1}$, whose elements are
triples $(g, \chi, \eta)\in \Gscr \ltimes \form{0}\ltimes
\form{1}$. The product of two such triples is given by
$$(g_1, \chi_1,\tau_1)(g_2,\chi_2,\tau_2)=
\left(g_1g_2, {\Ad}_{g_2^{-1}}\chi_1+ \chi_2, {\Ad}_{g_2^{-1}}
\tau_1+
\tau_2,
\right).\ref\product$$

The Lie algebra of $\Gaff$ is
$$\left(\form{0}\oplus_s
\form{0}\right)\oplus_s\form{1}$$ and the relevant commutator
is given by
$$\left[(\zeta_1,\chi_1,\tau_1),(\zeta_2,\chi_2,\tau_2)\right]=
\left([\zeta_1,\zeta_2], [\zeta_1,\chi_2]+[\chi_1,\zeta_2], [\zeta_1,\tau_2]+
[\tau_1,\zeta_2]\right).$$
\sezione{Tangent connections}

Now we discuss  the structure of the space of connections of
the principal $\T G$-bundle $\T P$.

For any manifold $M$ we can consider its double tangent bundle $\T\T M$
$$
\matrix{\T\T M&\riga{\T\pi_M }{20}&\T M\cr \mapdown{\pi_{\T M}}
&&\mapdown{\pi_M}\cr \T M&\riga{\pi_M}{20}&M.\cr }
$$
Let us consider a rectangle in $M$ centered at $x$, i.e., a $C^2$-map
$  \underline x\colon [a,b]\times [a',b']\mapsto M$, with $(0,0)\in (a,b)\times
(a',b')$,
$\underline x(0,0)=x$.
 We will denote by $s$ and $t$ the two variables and by a prime the derivative w.r.t.\
$s$, by a dot the derivative w.r.t.\ $t$.
A tangent vector in $\T\T M$ at $\dy{\left(\underline x(0,0),  \underline
  x'(0,0)\right)}$ can be represented as
$\dy{\left(\dot {\underline x}(0,0),{d\underline x'\over dt}(0,0)\right)}.$
When it is not otherwise specified,
the derivatives are understood to be
computed at  $(0,0)$.
Notice that two rectangles are meant to be
equivalent if they have the same first order derivatives
as well as  
the mixed derivatives of order 2. 
Everything that follows is defined on the equivalence classes.
Identifying double tangents with equivalence
classes of rectangles provides us also with a (non-canonical)
extension of the double tangent vector to the image of the rectangle.

In our notation we have, for a given
$[\underline x]\in \T\T M,$  $\T\pi_M
[\underline x]= \underline {\dot x}$ and $\pi_{\T M}
[\underline x]= \underline {x'}.$

There is a canonical involution
$
\alpha_M\colon \T\T M\mapsto \T\T M
$
given by
$$ \dy{
\alpha_M \big(x , \underline x', \dot {\underline x},{d \underline x' \over dt}\big)}
=
\dy{\left(x ,  \dot{\underline x} ,  \underline x' ,{d\dot{\underline x}\over ds}
\right)}.
\ref \alfa
$$
It is well known \quot\kob{S. Kobayashi, ``Theory of Connections",
Ann.\ Mat.\ Pura
Appl.\ {\bf 43}, 119--194
(1957);} that for any connection $A$ on $P$, one can induce a
connection
$\dota$
on $\T P$. If $\underline p $ is
any rectangle in $P$ centered at $p$, then $\dota$
 is, by definition, the
1-form on
$\T P$ with values in
${\frak g} \oplus_s {\frak g} $, given by
$$
\displaystyle{\dota_{p,\underline {  p'}}\left(  \underline{
\dot p}, {d\underline {  p'}\over dt} \right)  } =
\displaystyle{\left(A_p\left(\underline{ \dot p }\right),\dss
\left[A_{\underline p(s,0)
} \left( \dot  {\underline p}(s,0)\right)\right]\right).}
\ref\tangconn$$
Notice that the canonical
involution has been used in the definition of  the above connection.

A more general connection on $\T P$ can be obtained
by considering the evaluation map
$$
ev\colon \Ascr\times \T P\to {{\frak g}}
$$ $$
(A;p,X)\rsa A_p(X).
\ref\evaluation
$$
The tangent map  of \evaluation\
$$
\T ev\colon \T\Ascr\times \T\T P\to {{\frak g}}\oplus_s{{\frak g}},
\ref\evstar
$$
evaluated at $(A,\eta)\in \T\Ascr$ and composed with the canonical
involution $\alpha_P$ gives a connection on $
\T P$. In fact we
have the following
\theorem\evalconn{
For any $(A,\eta)\in \T\Ascr$, the 1-form $\omega(A,\eta)$ on
 $\T P$ given by
$$
\omega(A,\eta)([\underline p])\equiv
\T ev(A,\eta;\alpha_P[\underline p ])= \left(A_p\left(
 \underline {\dot p} \right) ,\quad \dss A_{\underline {  p}(s,0) } ( \underline
{\dot p}(s,0) ) +
\eta\left(
\underline{ \dot p} \right)\right)
\ref\Aeta
$$ defines a connection on $TP$.}
\proof\evalconn

The adjoint action of the group on its Lie algebra,
extends, in an obvious way, to Lie-algebra valued forms. So we have
a natural map $\ad\colon G\times \Omega^*(P,{\frak g})
\mapsto\Omega^*(P,{\frak g})
$ and
its derivative $\T\ad_G={\rm ad}_{\T G}$ (see \Adeq).
If we consider again the derivative in the
first variable of the right multiplication on $P$,
i.e., the map $\T_1 R: \T P \times G\to\T P$
and we denote by
$\Delta_G:G\to G\times G$ the
diagonal map in $G$ and by $p_{V}:V\times U\to V$ the projection
on $V$
for any pair of spaces $(V,U)$, then  the $\ad$-equivariance
of the connections is given by the following equation
$$ev\circ p_{\Ascr\times \T P}=ev\circ(\ad\times \T_1 R)
\circ (\Id\times \Delta_G):\Ascr\times
\T P\times G\to\frak g,\ref\equivariance$$
where the evaluation map is given by
\evaluation . The derivative of both sides of
\equivariance\  gives
$$\T ev \circ p_{\T \Ascr \times \T\T P}=
\T ev\circ ( {\rm ad}_{\T G}\times
\T_1 R)\circ\left (\Id \times \Delta_{\T G }\right).
\ref\equivariancestar
$$
The equivariance of \Aeta\ follows when we notice that
$$\T\T_1 R=\alpha_P \circ \T_1 \T R \circ \alpha_P:\T\T P \times \T G\to
\T\T P,$$  where
$\alpha_P$ is the canonical involution.
Finally we notice that the difference
$\omega(A,\eta)- \dota$ evaluated at $[\underline p]$ gives
$\eta(\underline {\dot p})$. But the projection $\T \pi_P$ applied
to the  fundamental
vector field in $(p,X)\in \T P$ determined  by
$(a,b)\in {\frak g}\oplus_s {\frak g}$ yields
$i(a)_p$, i.e., the fundamental
vector field in $P$ corresponding to $a$ evaluated at $p$.
The theorem follows from the fact that $\dota$ is a connection
and $\eta$ is a tensorial 1-form.
 \endproof

When we consider the  inclusion
$$\iota\colon P\incul \T P,\quad \iota(p)= (p,0),\ref\inclu$$
then the pullback of the connection $\omega(A,\eta)$
 via \inclu\ is simply given by
$$[\iota^*
\omega(A,\eta)]_p(X)=\left(A(X)_p, \eta(X)_p\right)\in {\frak g}\oplus_s
{\frak g},\quad
X\in \T_pP.\ref\pullback$$

Also we have the following
\theorem\connsubset{The inclusion
of $\T\Ascr$ into
the space of connections of $\T P$ is proper.}
\proof\connsubset

The bundle $
\T\Ascr$ can be identified with an affine space modelled on
$$\tform{1}\sim \ttform{1},$$ while the space of connections on $\T P$ can
be identified
with an affine space modelled on $\Tform{1}$. The theorem is proved
when we notice that the projection $M\to \T M$ determines a proper inclusion of
$\Omega^*(M)$ into $\Omega^*(\T M)$ and that the
  pair of projections $\T P\to P$, $\T G\to G$ determines:
\item{1.} a homomorphism of the $
\T G$-principal bundle $\T P\times {\frak g}
\times {\frak g}\to \adTP$ (where the left $\T G$-adjoint action on
${\frak g}
\times {\frak g}$ is considered)
onto the $G$-principal bundle
$P\times{\frak g}\times {\frak g}\to \tadP$ (where the left $G$-adjoint action
on ${\frak g}
\times {\frak g}$ is considered), and hence
\item{2.}  a proper inclusion of the space of sections
$\Gamma\left(\tadP\right)$ into
the space of sections $\Gamma\left(\adTP\right)$.
\endproof

Two final remarks are in order.

\item{A)} By iterating the procedure, we obtain an inclusion
$$\T^n\Ascr(P)\incul \Ascr(\T^nP)$$
where $\T^n$ denotes the $n$-iterated tangent bundle and $\Ascr(\bullet)$
denotes the space of connections of a given principal bundle.
\item{B)} When we consider only irreducible connections, then
$\dy{\Ascr\mapsto \Ascr/\Gscr}$ is a principal bundle and so is
$\dy{\T^n\Ascr\mapsto \T^n\Ascr/\T^n\Gscr}$.

\sezione{Topological field theories for tangent principal bundles}

Here we want to extend the topological field theory discussed in section
{\bf II} to the case
when the principal bundle is the {\it tangent bundle $\T P$}.
The purpose of this section is
to write the BRST equations for this case.

Here we will
consider only those special connections for $\T P$ that are
determined by elements of $\T\Ascr$, i.e., by pairs $(A,\eta)$ where
$A$ is a connection on $P$ and $\eta$ is any element of $\form{1}.$

In analogy with the
discussion of section {\bf II}
we will begin by including, among the admissible
symmetries, besides the tangent gauge group $\T\Gscr$,
the full translation group on the affine space $\T\Ascr$.

Namely, we want to consider the action of
the group $\T\Gscr_{\T}$
(tangent bundle of the group $\Gscr_{\T}$ defined in \gt\ )
on $\T\Ascr$. This
action is given by the
derivative of \transl
$$\eqalign{\T\Ascr \times \T\Gscr_{\T}&\to \T\Ascr,\cr
(A,\eta)(g,\rho,\xi,\tau)&\rsa (A^g+\rho,
\Adgm  \eta+d_{A^g}\xi+\tau).\cr} \ref\ttransl$$
Since
\transl\ is not free, neither is
\ttransl.

We have an obvious infinite-dimensional analogue of \evalconn:
\theorem\cc{Any pair $(c,{\hat c})$ where $c$ is a connection on
\gauorb\ and ${\hat c}$
is a tangent vector to the space of connections on
\gauorb, defines a connection $\omega(c,\hat c)$ on
$$\T\Ascr\mapsto {\T\Ascr\over \T\Gscr}.\ref\tgauorb $$}
\blank
Explicitly ${\hat c}$
is an assignment to each connection $A\in \Ascr$ of a map
${\hat c}_A\colon\form{1}\mapsto \form{0}$ with the property
of $\Gscr$-equivariance
$${\hat c}_{A^g}\left({\Ad}_{g^{-1}}\tau\right)
= {\Ad}_{g^{-1}}\left({\hat c}_A(\tau)\right),$$
and of tensoriality
$$\hbox{Im}\,\left(d_A\big|_{\form{0}}\right)\subset \ker ({\hat c}_A).$$
In physics ${\hat c}$ is an {\it infinitesimal variation of the gauge fixing}.

By using the same notation of the previous theorem, we have also the following
\theorem\ccbis{The pair $(c,\hat c)$ determines a connection on
the $\T G$ principal bundle $\T P\times \T\Ascr \to \T M \times \T\Ascr$ that is
$\ T\Gscr$-invariant, i.e., determines a connection on the principal
$ \T G$-bundle
$${{\T P\times \T\Ascr}\over{\T\Gscr}}\mapsto \T M\times {\T\Ascr\over \T\Gscr}.\ref
\tpetg$$}

If we consider a rectangle
$\left([\underline p],[\underline A]\right)$
on $\T P\times \T\Ascr$
centered at $(p,A)\in P\times \Ascr$, with $X=\underline{\dot p},$
and $\eta=\underline{\dot A}$,
then the explicit expression of such a
connection at $(p,X,A,\eta)\in \T P\times \T\Ascr$
is given by
$$\omega(A,\eta)([\underline p]) + \omega(c,\hat c)([\underline A]),
\ref\conntpetg$$
where the notation here is completely analogous to the one of
\evalconn.

A simple calculation shows
\theorem\ccter{
When we identify the double tangent bundle $\T\T\Ascr$ with $\Ascr\times
\form{1}^{\times 3}$, then the connection determined by
$(c,{\hat c})$ is a map
$$(A, \eta, \gamma, \sigma)\rsa \left(c_A(\gamma), \dss c_{A+s\eta}(\gamma)
+ {\hat c}_A(\gamma) + c_A(\sigma)\right).$$}

{}From now on we set
$$\xi_{A,\eta}(\gamma)\equiv \dss c_{A+s\eta}(\gamma) + {\hat c}_A(\gamma)
.\ref\defxi $$

Since space--time for us is $M$ and not $\T M$, we want
to avoid dealing with forms on $\T P$, as opposed to forms on $P$.

We consider then again the inclusion map \inclu\ $P\to \T P$
and establish the following
\convention\conv{All the forms on $\T P\times \T\Ascr$ are pulled
back to forms on $P\times \T\Ascr$.}
\blank
In the use of \conv\ we should be aware that, even though
\inclu\ is a morphism of principal
bundles, the pullback of a connection on $\T P \times \T\Ascr$ is
{\it not} (strictly speaking) a  connection on $P \times \T\Ascr$. It is
a $({\frak g}\oplus_s {\frak g})$-valued 1-form that would become
a connection only
if we decided to disregard the second copy of ${\frak g}$.

Moreover, when considering the decomposition $\T\Ascr\sim \Ascr\times
\form{1}$ we also assume the following
\convention\convbis{We will omit the components
of the forms on $P\times \T\Ascr$ which have degree higher than
0 as forms on $\form{1}$.}
\blank

When we assume both the previous conventions,
then the connection \conntpetg\ becomes the following $({\frak g}\oplus_s
{\frak g}$)-valued 1-form on $P\times \Ascr$, depending on  $\eta\in
\form{1}$

$$(A+c;\;\eta+\xi)_{p,A,\eta}(X,\gamma)\equiv \left(A_p(X) +c_A(\gamma);\;\eta_p(X)
+\xi_{A,\eta}(\gamma)\right), \quad X\in \T_pP, \gamma \in \T_A\Ascr
,\ref\connpetg$$
where \defxi\ has been used.

We will, in the future,  refer to \connpetg\ as
a ``connection" with an associate
``covariant exterior derivative,"
``curvature" and ``Bianchi identities;"
but, as has been explained above,
this will clearly be an abuse of language.

The covariant exterior derivative for $({\frak g}\oplus_s
{\frak g})$-valued forms on $P\times \Ascr$  is given by
$$d_{(A,\eta)} + \pm \delta_{(c,\xi)}.
\ref\covderiv$$
More precisely the
covariant exterior derivative applied to the pair of ${\frak g}$-valued
forms
$(\omega, \omega')$,
with degree $(k,k')$ as forms on $P$, is
given by
$$ \left(d_A \omega + (-1)^k \delta \omega+
(-1)^k[c,\omega],d_A\omega'
+(-1)^{k'}\delta\omega' +[\eta+(-1)^{k}\xi,\omega]+[(-1)^{k'}
c,\omega']\right) \ref\covder$$
where we used \commut.
The  curvature of \connpetg\ is a $({\frak g}\oplus_s
{\frak g})$-valued 2-form
on $P\times \Ascr$ which we can compute in an similar way  as
$$(F_A+\psi+\phi; \quad d_A\eta+ \tilde\psi+
\tilde\phi),\ref\curvtot$$
where $(F_A,d_A\eta)$, $(\psi,\tilde\psi)$ and $(\phi,\tilde\phi)$ are
respectively the $(2,0)$, the $(1,1)$ and the $(0,2)$ components.

We now compute explicitly the structure equation and the Bianchi identities.

First we obtain once again
the equations \brstpeg, \brstpegi\ and \brstpegii.

The structure equations give moreover
$$\tilde \psi=-\delta \eta + d_A\xi +[\eta,c],\ref\curv$$
$$\tilde \phi=\delta \xi + [c,\xi].\ref\curvi$$
Finally the Bianchi identities for the {\it second} component of the curvature
give
$$\delta \tilde \psi= -[\tilde \psi, c] + d_A \tilde \phi +[\eta,\phi]-
[\xi,\psi],
\quad \delta \tilde \phi =[\tilde \phi, c]+ [\phi,\xi].\ref\bianchi$$
So we have:
\theorem\fieldsuno{
The transformations for the set of fields $A,\eta,c,\xi,\psi,\phi,\tilde
\psi,\tilde \phi$ given by the components of the connection \connpetg\
and the curvature
\curvtot\ are as follows
$$\delta A= d_A c-\psi\ref\eqauno$$
$$\delta \eta=-\tilde \psi + d_A\xi +[\eta,c]\ref\eqadue$$
$$\delta F_A= -d_A\psi +[F_A,c]\ref\eqatre$$
$$\delta(d_A\eta)= -d_A\tilde\psi +[F_A,\xi]
+[d_A\eta,c] -[\psi,\eta]\ref\eqaquattro$$
$$\delta c = -{1\over 2} [c,c] +\phi\ref\eqacinque$$
$$\delta \xi=\tilde \phi - [c,\xi]\ref\eqasei$$
$$\delta \psi= d_A\phi -[\psi,c]\ref\eqasette$$
$$\delta \tilde \psi= -[\tilde \psi, c] + d_A \tilde \phi +[\eta,\phi]-
[\xi,\psi]\ref\eqaotto$$
$$\delta\phi=[\phi,c]\ref\eqanove$$
$$\delta \tilde \phi =[\tilde \phi, c]+ [\phi,\xi].\ref\eqadieci$$}

The transformation laws  in \fieldsuno\ show how to overcome
the problem of the lack of freedom of \ttransl. As before the key element
has been the introduction of
a (equivariant) coboundary operator relevant to a principal
bundle (in this case \tpetg).

\sezione{Restrictions to the orbits}

Two possible fiber imbeddings can be considered:
$$j_{A}\colon \Gscr\incul\Ascr,\quad j_A(g)= A^g, \ref\ja$$
$$j_{(A,\eta)}\colon \T\Gscr\incul \T\Ascr,\quad j_{(A,\eta)}(g,\chi)= \left(A^g,
{\Ad}_{g^{-1}}(\eta) + d_{A^g}\chi\right).\ref\jaeta$$

First we can pull back the  bundle
$$P\times \T\Ascr\mapsto M \times
\T\Ascr \ref\pta$$ to
$$P \times \T\Gscr\mapsto M\times \T\Gscr$$ via \jaeta.

Then we pull back the 1-form $(c,\xi)$ (without changing notation).
This form then becomes
the Maurer--Cartan form on $\T\Gscr$, and the structure equations
and Bianchi identities become simply
$$\delta A= d_A c,\quad \delta c = -{1\over 2} [c,c],\quad \delta
\xi=-[c,\xi], \quad \delta \eta= d_A\xi +[\eta,c].
\ref\brsi $$
This is the generalization to the tangent bundle of the ordinary BRST
equations \brs.

Starting from \ja\
we   obtain another restriction
that will allow us to have a BRST symmetry that includes
the 2-form field $B$ considered in 4-dimensional
$BF$ theories.

We begin by noticing that the group $\Gaff$ considered in section
{\bf III} is the right symmetry group for the action functional
\bfymeta. Here the action of
$\Gaff$  on triples $(A,\eta, B)$
with $(A,\eta)\in \T\Ascr$ and $B\in\form{2}$ is defined as
follows\foot{Notice that this action of $\Gaff$ on $\T\Ascr$ does {\it not}
coincide with the restriction of the action of $\Gscr_{\T P}$ on $\Ascr(\T P).$}

$$(A,\eta)\cdot(g,\zeta,\tau)= \left( A^g, \tau + {\Ad}_{g^{-1}} \eta +
d_{A^g} \zeta\right),\ref\gaffaction $$
$$B\cdot(g,\zeta,\tau)=
 \Adgm B + d_{A^g} \tau + d^2_{A^g} \zeta.\ref\gaffactionb $$

The group $\Gaff$ is just a subgroup  of $\T\Gscr_{\T}$
since we have
$$\Gaff\equiv j^*(\T\Gscr_{\T}),$$
where $j$ is the inclusion
$j\colon \Gscr\incul\Gscr_{\T}.$
The action \gaffaction\ is the corresponding restriction of \ttransl.

Requiring the invariance under
the group of affine gauge transformations is tantamount
to considering a {\it ``semi-topological" field theory where both
the invariance under $\T\Gscr$ and the independency of the choice of the
tangent vector $\eta\in \T_A\Ascr$ is considered.}

To turn around the problem of the non freedom of $\gaffaction$
we proceed along the same lines of the previous section.

First we consider the
pulled-back  bundle $j_A^*\T\Ascr$ and
pull back to it the 1-form  $(c,\xi)$ (again without changing notation).

Now, as in \brsi,
$c$ becomes
the Maurer--Cartan form on $\Gscr$, but $\xi$, differently
from \brsi, does not become any more the second component of
the Maurer--Cartan form on $\T\Gscr.$ The form $\xi$
is defined as in \defxi, provided that
the connection on $P$  belongs to the $\Gscr$-orbit passing through $A$.

The corresponding pulled-back ``connection" on the bundle
$$P\times j_A^*\T\Ascr\mapsto M\times j_A^*\T\Ascr,\ref
\jpetg$$
is written again as $$(A+c,\eta +\xi)\ref\connsemitot$$
and the corresponding ``curvature" is
$$(F_A, d_A\eta +\tilde \psi +\tilde \phi).\ref\curvsemitot$$

The Bianchi identities become
$$[d_{(A,\eta)} \pm \delta_{(c,\xi)}](F_A,d_A\eta + \tilde \psi +\tilde \phi)
=0$$
and altogether we obtain (see  \bfseven):

\theorem\fieldsdue{
The transformation laws for the set of fields $A,\eta,c,\xi,\tilde
\psi,\tilde \phi$ given by the components of the connection \connsemitot\
and of the curvature
\curvsemitot\ are as follows:
$$\delta A= d_A c \ref\eqbuno$$
$$\delta \eta=-\tilde \psi + d_A\xi +[\eta,c]\ref\eqbdue$$
$$\delta F_A=[F_A,c]\ref\eqbtre$$
$$\delta(d_A\eta)= -d_A\tilde\psi + [F_A,\xi]
+[d_A\eta,c]\ref\eqbquattro$$
$$\delta c = -{1\over 2} [c,c]\ref\eqbcinque$$
$$\delta \xi=\tilde \phi - [c,\xi]\ref\eqbsei$$
$$\delta \tilde \psi= -[\tilde \psi, c] + d_A \tilde \phi\ref\eqbsette$$
$$\delta \tilde \phi =[\tilde \phi, c].\ref\eqbotto$$}

By comparing \eqbuno, \eqbdue\ with \gaffaction\ we see that also  this
last BRST operator $\delta$ allowed us to
overcome the problem of the lack of freedom of \gaffaction.
We have still to find how to read  in our BRST complex the transformations
of the field $B$ corresponding to the action
\gaffactionb. This will accomplished in the next section.

\sezione{A BRST complex that includes 2-forms}

Now we will show that  the BRST operator $\delta$
can consistently be
extended to an operator $s$ satisfying
$s^2=0$ and that
a double complex with operators $(d,s)$
can be constructed  with the
following properties:
\item{1.} $s$ acts on the space $\form{2}$;
\item{2.} the gauge-equivariance is preserved, and
\item{3.} the equations considered in \fieldsdue\ are
preserved.

We use the synthetic notation
$\Bscr\triangleq \form{2}$
and consider the  tangent bundle
$\T\Bscr\sim \form{2}\times \form{2}.$

The group $\Gscr$ acts on $\Ascr\times \Bscr$ yielding a $\Gscr$-principal
bundle. Moreover, the tangent map to the above action gives $\T\Ascr \times
\T\Bscr$ the structure of a $\T\Gscr$-principal bundle.

Explicitly this last action
is given by
$$(A,\eta;C, E)\cdot (g,\zeta)=$$
$$\left( A^g, {\Ad}_{g^{-1}}\eta
+d_{A^g} \zeta; {\Ad}_{g^{-1}} C, {\Ad}_{g^{-1}}E +
[{\Ad}_{g^{-1}}C,\zeta]
\right),\ref\gaugeactionab.$$

It is evident that
the projection $\Ascr\times \Bscr\mapsto \Ascr$ is a morphism
of $\Gscr$-bundles. Therefore,
the connection $c$ on \gauorb\ can also be considered as
a connection
on $\Ascr\times \Bscr$, and the  connection $(c,\xi)$ on \tgauorb\
is also a connection on $\T\Ascr\times \T\Bscr$.
\blank
Moreover, $(A+c, \eta+ \xi)$ is a (pulled-back)
$({\frak g}\oplus_s {\frak g})$-valued  connection on the bundle
$${P\times \T\Ascr\times \T\Bscr}\mapsto M\times \T\Ascr\times
\T\Bscr.\ref\petgab$$

The corresponding covariant exterior derivative is given by \covderiv.
Forms on $P\times \T\Ascr\times \T\Bscr$ will be characterized by
three indices $(m,s,p)$ which represent the degree with respect to the
three spaces $P$, $\T\Ascr$, $\T\Bscr.$ The integer $s$
is, as before,  called the {\it ghost number}.

The pair $(C,E)\in \T\Bscr$ is a $({\frak g}\oplus_s {\frak g})$-valued
$(2,0,0)$-form that
is independent of $\T\Ascr.$

We assume  \conv\ and \convbis, and  also the following
\convention\convter{
We will omit the components
of the forms on $P\times \T\Ascr\times \T\Bscr$
which have degree higher than
0 as forms on $\T\Bscr$.}
\blank
Under the action of
the covariant exterior derivative \covderiv, the pair
$(C,E)$ is transformed into
$$\left(d_AC + [c,C],d_A E +[\eta,C]+[c,E] +
[\xi,C]
\right),\ref\omegabi$$
owing to the fact that $(C,E)$ does not
depend on the space $\Ascr$.

\definizione\boh{We define $s$ to be equal to $\delta$ when computed
on forms over $P\times \T\Ascr$;
moreover, we set for $(C,E)\in \T\Bscr.$
$$s(C,E)\triangleq ([C,c], [C,\xi] +[E,c]).\ref\deltaeq $$}
\blank

We have hence set $s(C,E)$ to be equal, up to a sign, to
$\delta_{(c,\xi)}(C,E).$ This takes care of requirement 2
(preservation of gauge-equivariance).

As a consequence we have
$$\delta_{(c,\xi)}^2(C,E)=[(\phi,\tilde \phi),(C,E)]=([\phi,C],[\tilde \phi,C]
+[\phi,E]),$$
where $(\phi,\tilde \phi)$ is the curvature of the connection $(c,\xi)$
on the
$\T\Gscr$ bundle $\T\Ascr$. Here \conv\ and \convbis\ are always assumed.

In conclusion we have:
\theorem\bbrs{Let us assume \boh.
The operator $s$ is nilpotent
in the following cases:
\item{1.} always in the case of equations \brsi;
\item{2.} never in the case of the equations of \fieldsuno;
\item{3.} only when pairs $(0,E)\in \T_0\Bscr$ are considered
in the case of the equations of \fieldsdue.}

\remark\remuno{The reason why,
when we consider the bundle \petgab, we have
non-trivial solutions for the compatibility problem of \bbrs,
is the fact that $\{1\}\times {\frak g}\triangleleft G\ltimes {\frak g}$
in an abelian normal subgroup.}
\blank

We now translate the field $E$ and define
$$B\equiv E+d_A\eta.\ref\bnew$$
By combining \eqbquattro\ with \deltaeq\
we find the following transformation property of $B$ (see \bfseven\ and
\quot\fucito{F.Fucito, M.Martellini,
A.Tanzini, M.Zeni, unpublished;})
$$sB = -d_A\tilde \psi +[F_A,\xi] + [B,c].\ref\bnewuno$$
Equation \bnewuno\ may
be added to  equations \eqbuno\ $\lora$ \eqbotto, namely,
$$sA= d_A c,\quad
s\eta=-\tilde \psi + d_A\xi +[\eta,c],$$
$$s c = -{1\over 2} [c,c],\quad s \xi=\tilde \phi - [c,\xi],$$
$$s\tilde\psi= -[\tilde \psi, c] + d_A \tilde\phi ,\quad
s \tilde \phi =[\tilde \phi, c].$$

While $E$ is
an element of $\T_0\Bscr$, $B$ is naturally
an element of $\T_{F_A}\Bscr.$
In particular,
triples $(A,\eta, B)$
are elements of
the  pulled-back bundle
$$K^* (\T\Ascr \times \T\Bscr),$$
where $K$ is the curvature map
$$K:\Ascr\mapsto\Ascr\times \Bscr,\quad K(A)\triangleq(A,F_A). \ref\curvmap$$

As a final remark, notice that
once again
we
have overcome the problem of the lack of freedom of \gaffactionb.

\sezione{Infinite BRST symmetries}
The procedure of the previous two sections can be iterated, so as
to incorporate
an arbitrary number of fields. More precisely, given any integer $n$
we can construct field equations, similar to those considered in the previous
section, depending on a connection $A$,
on $2^n-1$ fields $\gamma_i\in \form{1}$, on $2^n$ fields
$\eta_i\in \form{1},$ on $2^n$ fields $B_i\in\form{2}$, and on the relevant
ghost-fields.
\blank
A binary notation turns out to be
helpful to describe iterated tangent spaces.
In fact  given a manifold $M$ we   will
describe an element ${\bf X}\in \T^nM$  by a $2^n$-tuple as follows
$${\bf X} =\left ( X_{{\underbrace{0\dots 00}_{n-{\rm
times}}}},X_{{\underbrace{0\dots 01}_{n-{\rm times}}}},X_{{\underbrace{0\dots
10}_{n-{\rm times}}}},\dots,
X_{{\underbrace{1\dots
11}_{n-{\rm times}}}}\right).$$
For instance in the case $n=3$, $X_{0 0 0}$ denotes a point of $M$,
$\left(X_{000}, X_{0 0 1}\right)$ denotes a point in $\T M$,
$\left(X_{000}, X_{0 0 1},X_{010}, X_{011}\right)$
denotes a point in $\T\T M$ whose tangent vector is represented by
$\left(X_{100}, X_{1 0 1},X_{110}, X_{111}\right).$

Elements in
$Lie(\T^nG)$ can analogously be   represented  as $2^n$-tuples
$\bxi=\left(\xi_{i_1,\dots,i_n}\right)$ where each
index $i_p$ can be either $0$ or $1$,
$\forall p\in \{1,\dots, n\}$.

\blank
We now want to describe   the bracket in  $Lie(\T^nG)$ induced by the
bracket in
$\frak g$.
We have
\theorem\bractg{
The Lie bracket of $\bxi$ and $\beeta$ in $Lie(\T^nG)$ is given by:
$$[\bxi,\beeta]_{i_1\cdots i_n}=\sum_{
j_i+k_i=i_i}[\xi_{j_1\cdots j_n},\eta_{k_1\cdots k_n}]\ref\bracket$$
where  the previous notation is understood.}
\proof \bractg

We will  prove \bracket\ by induction on $n$. Formula
\bracket\ holds for $n=0$, i.e., for
$\frak g$. We assume now that
it is true for $n$. We can represent any element $\bxi$ of
$Lie(\T^{n+1}G)$ as $(\xi_{0j_1\cdots j_n},\xi_{1j_1\cdots j_n})$
where the first
component represents the base point in $Lie(\T^{n}G)$, while
the
second component represents the relevant    tangent  vector.
We can therefore use the expression of the bracket for
$Lie(\T G)\sim {\frak g}\oplus_s {\frak g}$ thus getting
$$[(\xi_{0j_1\dots j_n},\xi_{1j_1\dots j_n}),(\eta_{0k_1\cdots
k_n},\eta_{1k_1\dots k_n})]=\bigg([ \xi_{0j_1\dots j_n},\eta_{0k_1\dots
k_n}],$$ $$[\xi_{0j_1\dots j_n}, \eta_{1k_1\dots k_n})]+
[\xi_{1j_1\dots j_n},\eta_{0k_1\dots k_n})] \bigg)
$$ Therefore,
$$[\bxi,\beeta]_{i_0i_1\dots i_{n }} =\sum_{ j_i+k_i=i_i}[\xi_{j_0j_1\dots
j_n},\eta_{k_0k_1\dots k_n}] $$
\endproof
We now take into account the iterated gauge group
$\T^n \Gscr.$

We can consider the
Maurer--Cartan form for $\T^n\Gscr$. This is a form with values in
$Lie(\T^{n}\Gscr)$, so it can be represented by a vector
${\bf c}$ with $2^n$ components
each having ghost number one.

We set $c_0\equiv c_{0\dots 0}$, and $ c_I\equiv c_{i_1\dots i_n}$
for $I=\{i_1,\dots,i_n\}\neq \{0\dots 0\}$, and define
$I+J\triangleq \{i_1+j_1,\dots,i_n+j_n\}$ ({\it not} modulo 2).

The  {\it Maurer--Cartan}
equations are as follows:
$$\delta c_0 =-{1\over 2} [c_0,c_0], \ref\infuno$$
$$\delta c_I =-{1\over 2}\sum_{J+K=I} [c_J,c_{K}] .\ref\infdue$$

The group $\T^n \Gscr$ acts (freely)
on the iterated tangent space $\T^n \Ascr$. We denote the
corresponding fiber imbedding by
$$j^{(n)}_{A,\bgamma}\colon \T^n\Gscr\to \T^n\Ascr,$$
 where
$(A,{\bgamma})$ represents an element of $\T^n\Ascr$ with
${\bgamma}\in \form{1}^{\times 2^n-1}$.

In this section we consistently apply \conv, \convbis\ and \convter.
The equations describing the infinitesimal
action of $\T^n \Gscr$ and corresponding to \brsi\
are as follows:
$$\delta A= d_A c_0\ref\inftre$$
$$\delta \gamma_I = d_A c_I +\sum_{J+K=I }  [\gamma_J,
c_{K}] .\ref\infquattro$$

In analogy with \jpetg\ we apply the tangent functor
once more and consider the bundle
$$P\times\left(j^{(n)}_{\hbox{$A,{\bgamma}$}}\right)^* \T(\T^n\Ascr)\to
M\times\left(j^{(n)}_{\hbox{$A, \bgamma $}}\right)^* \T(\T^n\Ascr).\ref\infbundle$$
We have a vector ${\beeta}$ with $2^n$ components in
$\form{1}$ with ghost number zero
and a vector ${\bxi}$ with $2^n$ components in $\form{0}$ with ghost number
one.
In analogy with   \connsemitot\
on \infbundle\ we have a (pulled-back) connection\foot{We have to keep
in mind that each of the four vectors
$(A,{\bgamma}); {\beeta}; {\bf c}; {\bxi}$ is a form
on $P\times \T^n \Ascr$ with values in $Lie(\T^{n}G)$.}
$$\left( (A,{\bgamma})+{\bf c}, {\beeta}+{\bxi}\right)\ref\infconn$$
with values in $Lie(\T^{n+1}G).$

The curvature of \infconn\ is given by
$$ \left(F_A+ d_A{\bgamma}+{1\over 2}[\bgamma,\bgamma], d_A{\beeta}+
[{\bgamma,\beeta}]+ { \tilde \bpsi} +{  \tilde
\bphi}\right),\ref\infcurv$$
where the $I$-th component of the vector $[{\bgamma,\beeta}]$
is defined to be $\sum_{J+K=I}[\gamma_J, \eta_{K}],$
the vector
${  \tilde \bpsi}$ has $2^n$ components in $\form{1}$ and ghost number one,
while the vector ${  \tilde
\bphi}$ has $2^n$ components respectively in $\form{0}$ and ghost number two.
\blank
The results of the previous section,
suggest us to introduce a vector ${\bf B}$ with $2^n$ components in $\form{2}$
and ghost number zero.

By performing calculations similar to those carried on in
the proof of
\fieldsdue\
and  \bnewuno,
we obtain the following
\theorem\fieldsinf{
The transformations for the set of fields $A,{\bgamma},
{\beeta},{\bf c},{\bxi},{  \tilde
\bpsi},{ \tilde \bphi},{\bf B}$ are given by
\infuno, \infdue, \inftre, \infquattro\
and
$$\delta \eta_I=-\tilde \psi_I +
d_A\xi_I +\sum_{J+K=I}[\gamma_J,\xi_K] +\sum_{J+K=I}
[\eta_J,c_{K}]\ref\infcinque$$
$$\delta B_I= -d_A\tilde\psi_I -\sum_{J+K=I}[\gamma_J,\tilde\psi_{K}]
+ [F_A,\xi_I] +\sum_{J+K=I} [d_A\gamma_J,\xi_{K}]
+\sum_{J+K=I} [B
_J,c_{K}]\ref\infsei$$
$$\delta \xi_I=\tilde \phi_I - \sum_{J+K=I} [c_J,\xi_{K}]\ref\infsette$$
$$\delta \tilde \psi_I= -\sum_{J+K=I}[\tilde \psi_J, c_{K}] +
d_A \tilde \phi_I +\sum_{J+K=I}  [\gamma_J, \tilde\phi_{K}]\ref\infotto$$
$$\delta \tilde \phi_I =\sum_{J+K=I}[\tilde \phi_J, c_{K}].\ref\infnove$$}

\sezione{The Batalin Vilkovisky alternative: topological $BF$
theories}

In sections {\bf VI} and {\bf VII} we have considered the BRST structure
of a field theory with fields $(A,\eta)\in \T\Ascr$ and $B\in \Bscr$
that is gauge invariant and
can be considered {\it topological} only
as far as the translations of the field $\eta$ are concerned.
Equivalently such a theory admits, as its basic symmetry group, the group
$\Gaff.$

For a full topological $BF$ field theory, where
translations in the space $\Ascr$ have to be included
as symmetries, one should consider instead the action
of the group $\Gscr_{\T}$ on $\Ascr$ given by \transl\ and extended to
$\Ascr \times \Bscr$ as follows:
$$\left(A,B\right)
\cdot (g,\tau)=
\left(A^g + \tau, {\Ad}_{g^{-1}}B -
d_{A^g} \tau -{1\over 2}[\tau,\tau]\right).\ref\translb$$

Notice that \translb\ implies that the action of $\Gscr_{\T}$
on $B+F_A$ is simply the adjoint action of $\Gscr$.

An example of such a theory is given by the action functional \bftop\ when
$t=1$, i.e., by
$$S_{BF,1}= \int_M {\rm Tr}\left(F_A\wedge B + {1\over 2} B\wedge B\right).\ref
\bftopuno$$

Again the action \translb\ is not free,
but now this problem cannot be solved by
considering the coboundary (BRST) operator of a principal bundle.

In fact one would have  to extend the coboundary operator
$\delta$ defined on \peg\ to a nilpotent operator
$s$ acting on $B,$ while preserving the gauge equivariance.

This would mean to consider a covariant derivative acting on $\Bscr$
corresponding to a connection $c$ on $\Ascr$. The requirement of
the nilpotency of such an operator would then be equivalent to
the requirement that,
for any $B\in \Bscr$, the following condition should hold:
$$[\phi,B]=0,\ref\curvflat$$
where $\phi$ is the curvature of $c$.

But we know that condition \curvflat\ is not satisfied.

The so called Batalin--Vilkovisky mechanism \quot\batalin
{I. A. Batalin and G. A. Vilkovisky: ``Relativistic
S-Matrix of Dynamical Systems with Boson and Fermion Constraints",
\pl{69 B}, 309--312 (1977);
E. S. Fradkin and T. E. Fradkina: ``Quantization of Relativistic
Systems with Boson and Fermion First- and Second-Class Constraints",
\pl{72 B}, 343--348 (1978);} now comes into the picture by
providing us with a nilpotent coboundary operator that has the same
properties of a coboundary operator for a principal bundle.

The price we have to pay is that we have to
introduce fields with negative ghost-number
or, equivalently, that we have to consider fields that depend explicitly
on the (functional integral with respect to
the) action that we are considering.

The functional integral
{\it formally} provides us  with a
volume form on the space of fields $$\omega\equiv
\exp (-S) *1\equiv \exp(-S) \prod_x \Dscr A(x) \; \Dscr B(x),\ref\volume$$
where $S$ is the action functional \bftopuno.
The understanding that such a volume form exists is at the heart of
the Batalin--Vilkovisky mechanism, which is the same to say
that the Batalin Vilkovisky
mechanism is a device that allows {\it integration by parts in
field theories}.

Let us denote by ${\frak F}$ the ``manifold" given
by $\prod _j\prod_{x\in M} \Fscr_j (x)$ where $\{\Fscr_j\}$
is the set of
all fields of the theory (e.g., $A^b,B^c,c^a,\psi^d,\phi^e$,
with $c,\psi,\phi$ being defined  as
in section {\bf II}) and the indices $a,b,c,d,e$ labelling a basis for the
Lie algebra ${\frak g}$. We work here in local
coordinates  as far as the bundle $P(M,G)$ is concerned.

We denote by the symbol $B^*$ the (${\frak g}$-valued)
contraction with respect to the
vector field corresponding to the coordinate
$B^a(x),\; x\in M$ of the ``manifold" ${\frak F}$.
Namely, we set
$$(B^*)^a\triangleq \dy{\partial \over \partial B^a(x)}.\ref\bstar$$
Being the ghost number defined as the degree of a form on ${\frak F}$
(and particularly on the space of connections $\Ascr$),
a vector field on ${\frak F}$ will have ghost number $-1.$
More generally, given a field $\sigma$ with form degree $k$ (on $M$)
and ghost number $s$, its antifield $\sigma^*$ will have
form degree $4-k$ and ghost number $-s-1$.
The duality between the $\sigma$ and $\sigma^*$ will be given
by the following condition:
$$\{(\sigma^*)^a(x), \sigma^b(y)\} = \delta^a_b\,\delta(x-y),$$
where the Dirac volume form $\delta$ on $M$ has been used,
and $\{\bullet,\bullet\}$ denotes the Poisson bracket for the cotangent
bundle $\T^*{\frak F},$ which
is mapped onto the tangent bundle
via the metric (which gives the $*$-operator).

We now denote by $\delta$ the exterior derivative on ${\frak F}$.
By definition of $B^*$ we have
$$\delta\left(B^* (*1)\right)=0$$
and so
$$\delta( B^* \omega) =  (-B(x)-F_A(x)) \omega; $$
in other words we have the formal equation
$$\delta (B^*) = -B -F_A.$$

If we require equivarance, i.e.,
if we want to take
into account the action of the group of gauge transformations
$\Gscr$,
then we have to replace the exterior derivative with the covariant exterior
derivative  which we denote again by the symbol $s$.
So we have
$$s(B^*) = -B-F_A - [B^*,c],$$
and the ensuing transformation
$$s B= [B,c] + d_A \psi - [B^*,\phi].\ref\sbbv$$
It is interesting to compare \sbbv\ with \bnewuno.
The main difference between the two equations is that,
in the second case,  the
operator $s$ applied to $B$ produces {\it only fields},
while it produces {\it both fields and antifields} in the first case.
In the BV mechanism, antifields are added so to render the operator
$s$ nilpotent, i.e., so as to formally reconstruct a BRST operator.
This is the case of the topological $BF$ theory.
On the contrary, in the BFYM theory, antifields are unnecessary since the
BRST operator
is nilpotent {\it per se.}

{}From the previous equations we have
$$\int { \left[(B+F_A)\, \omega \right]\Dscr A\, \Dscr B\,}=0,$$
and in particular,
$${{\partial \exp S}\over {\partial B}} \approx 0,$$
$$ F_A \exp (-S) = \exp \left(-{1\over 2}\int_M{\tr(B\wedge B)}\right)
B^* \left [\exp \left(-\int_M\tr(B\wedge F_A)\right)\right]
\approx $$ $$- \exp \left(-\int_M{\tr(B\wedge F_A)}\right)
B^* \left[\exp \left(-{1\over 2}\int_M{\tr(B\wedge B)}\right)\right]
=-B \exp (-S),$$
where $\approx$ means up to a total derivative.
When the action functional is the topological $BF$ action functional
\bftopuno, then $B^*(x)$ is
equivalent to the transformation (concentrated at $x$)
$$F_A(x)\rsa - B(x).\ref\bvsf$$

We now include the fields (with negative ghost number)
$A^*,c^*,\psi^*,\phi^*$ conjugate respectively to
$A,c,\psi,\phi$.
The BV equations are written in
the following double-complex\vskip20pt
\hskip-12pt\vbox{\offinterlineskip  \def\tablerule{\noalign{\hrule}}
\halign to13.7cm{\strut#&\vrule#\tabskip=0.5em plus 2em& \hfil#&
\vrule#& \hfil#\hfil&\vrule#& \hfil#& \vrule#& \hfil#\hfil&\vrule#&
\hfil#& \vrule#&    \hfil#&\vrule#\tabskip=0pt\cr\tablerule
4&&&&&&&&$  \scriptstyle{s c^*= [c^*,c]-d_AA^*+ }$&&$
\scriptstyle{c^*=s\phi^*+[\phi^*,c]
}$&&$\phi^*\hskip-2pt$&\cr  &&&&&&&&$
\scriptstyle{[B^*,B]-[\psi^*,\psi]-[\phi^*,\phi] }$&&$
\scriptstyle{ +d_A\psi^*+{1\over 2}[B^*, B^*]}$&&&\cr\tablerule 3&&&&&&
$\scriptstyle{s
A^*+[A^*,c]=-d_AB}$&&$\scriptstyle{A^*=-s\psi^*+
}$&&$\psi^*$\hfil&&&\cr  &&&&&& $\scriptstyle{
-[B^*,\psi]+[\psi^*,\phi]}$&&$\scriptstyle{
+[\psi^*,c]+d_AB^*}$&& &&&\cr\tablerule 2&&&& $\scriptstyle{s
B+
[c,B]= }$&& $\scriptstyle{-B=s B^* }$&&$B^*$&&&&&\cr
&&&&$\scriptstyle{=d_A\psi+[B^*,\phi]}$&& $\scriptstyle{
+[B^*,c]+F_A}$&& &&&&&\cr\tablerule
1&&$\scriptstyle{-s\psi-[c,\psi]
}$&&$\scriptstyle{\psi=d_Ac-s A}$&& A&&&&&&&\cr
&&$\scriptstyle{ =-d_A\phi}$&& && &&&&&&&\cr\tablerule
0&&$\scriptstyle{\phi=s c }$&&$c$ &{}&& &&&&&&\cr
&&$\scriptstyle{ +{1\over 2}[c,c]}$&&  &{}&& &&&&&&\cr\tablerule  &
 &  2 & & 1 && 0 &&-1&&-2&&-3&\cr\tablerule &&\multispan{10}\hfil
$\liga{s}{70}  $\hfil&&\cr \tablerule\cr}} \vskip20pt
The degrees corresponding to rows and columns in the above diagrams
are respectively the form degree and the ghost number.
Notice that the form $$c+A + B^* + \psi^* + \phi^*$$ is formally
a ``connection"
with ``curvature"
$$\phi + \psi -B + A^* + c^*.$$
So the field $B^*$ allows the replacement,
in the BV complex, of $F_A$ with $-B$ in the structure
equations. This is consistent with transformation \bvsf\ and with
integration by parts in the functional integral.

At the critical point $B=-F_A$, or equivalently,
if we  set $B^*=0$, we obtain, from the previous complex,
the equations \brstpegi\ and \brstpegii.

\blank
In conclusion
topological field theories of the cohomological type are described
by the structure equations and the Bianchi identities of a suitable bundle
whose cohomology is some kind of equivariant cohomology.
In topological field theories which are not of cohomological type,
the above description is not possible, but the BV
mechanism, with the introduction of fields with negative ghost number,
allows us to recover, formally {\it off-shell},
the algebraic structure of equivariant cohomology.

\sezione{Gauge fixing and orbits}

In this section we would like to report briefly on the gauge-fixing problem
for the action functional \bfymeta.

The results of sections {\bf VI} and {\bf VII} imply that we have an exact
BRST symmetry when we consider the space of fields
$K^*(\T\Ascr \times \T\Bscr)$ and the symmetry group
$\T\Gscr$.

The action \bfymeta\ is invariant under $\Gaff$ and hence under its
subgroup $\T\Gscr$.

So the action \bfymeta\ is naturally defined on the
space
$${K^*(\T\Ascr
\times \T\Bscr)\over \T\Gscr}=K^*\left({\T\Ascr \times \T\Bscr\over
\T\Gscr}\right);$$
the last identity is a consequence of the fact that the curvature map
$K$ \curvmap\ commutes with the action of $\Gscr$ and hence descends to a map
$$K\colon{\Ascr\over\Gscr}\to {{\Ascr\times\Bscr}\over \Gscr}.$$

We notice that the following bundles are isomorphic:
$${K^*(\T\Ascr\times \T\Bscr)\over \T\Gscr}\simeq {K^*(\H\Ascr\times
\T\Bscr)\over\Gscr},\ref\torbit$$
the isomorphism being induced by
the linear map
$$[A,\eta,B]_{\T\Gscr}\rightsquigarrow
[A,\eta^\H,B-d_A\eta^\V]_\Gscr.\ref\map$$
Here $\H\Ascr$ denotes the horizontal bundle defined by a connection $c$
on $\Ascr$ and the superscript $\V$ and $\H$ denote respectively
the vertical and the horizontal
projections.

In order to show that this map is well-defined on the equivalence classes
we compute  $$\left(A^g,Ad_{g^{-1}}\eta+d_A\xi,Ad_{g^{-1}}B+[F_{A^g},\xi]
\right)\rsa$$ $$
(A^g,(Ad_{g^{-1}}\eta)^\H,Ad_{g^{-1}}B+[F_{A^g},\xi]-d_{A^g}\left
((Ad_{g^{-1}}\eta)^\V+d_{A^g}\xi\right)=$$ $$
\left(A^g,(Ad_{g^{-1}}\eta)^\H,Ad_{g^{-1}}B-
Ad_{g^{-1}}\left(d_{A}\eta^\V\right)\right).$$
It is immediate to check that \map\ is injective and surjective.

The action functional \bfymeta\ is also
invariant under the subgroup of $\Gaff$ given by the translations by
$\form{1}$
$$(A,\eta,B)\rsa (A,\eta+\tau,B+d_A\tau)\quad \tau\in \form{1}.\ref\trans$$

When our space of fields
is  represented by $\dy{K^*(\H\Ascr\times
\T\Bscr)\over\Gscr}$,  then we have to consider triples
$(A,\eta,B)$ with $\eta\in \H_A\Ascr.$ The action
\trans\ now becomes:
$$(A,\eta,B)\rsa (A,\eta+\tau^H,B+d_A\tau^H)\quad \tau\in \form{1}.\ref
\transone$$
The above action is of course not free, but the space
of orbits is still a manifold, namely it is
$${{\Ascr\times \Bscr}\over \Gscr}\sim {{K_2^*\T\Bscr}
\over \Gscr}.\ref\orbone$$

Different gauges are possible to describe the above orbit spaces.
Let us clarify first what we mean by ``choice of the gauge" in the present
framework.

We have the vector bundle $$\Vscr\triangleq
K^*(\H\Ascr\times \T\Bscr)\to \Ascr
\ref\vectbdlone$$
which is isomorphic to the vector bundle $$(\H\Ascr\times \Bscr) \to
\Ascr.$$ The $\Gscr$-invariant
fiber metric in $\H\Ascr$ and in $\Bscr$
induces a $\Gscr$-invariant fiber metric in $\Vscr.$

Fixing a gauge means finding a vector bundle $\Wscr$
so that
we have a $\Gscr$-equivariant splitting
$$\Vscr=\H\Ascr \oplus \Wscr.$$

The first obvious choice is to
choose $$\Wscr=\Ascr\times \Bscr.\ref\trivgauge$$
In other words the gauge condition reads:
$\eta=0,$ meaning that the elements of $\Wscr$
have zero projection on $\H\Ascr.$
We refer to the above gauge as to the {\it trivial gauge}.

In order to discuss
the second choice of the gauge, we
restrict $A$ to belong to the open subset of $\Ascr$, where
the dimension of the space $\harm{1}$
of harmonic 1-forms has the lowest dimension
(possibly zero). Let us denote this open set by the symbol
$\hat\Ascr$.

The symmetry \transone\ indicates that
we have at our disposal a second gauge fixing:
namely we can impose on $B$
the following condition $\langle B,d_A\eta\rangle =0$,
i.e., $\langle d_A^*B,\eta \rangle=0$, $\forall \eta\in
\H\Ascr$.
Here $\langle \bullet, \bullet \rangle$
denotes the inner product in $\form{*}.$

Our second gauge fixing is then a condition on both $\eta$ and
$B\in \Bscr$; namely, we require $\eta\in \H_A\hat\Ascr$ and
$B\in\hat\Bscr$, where $\hat\Bscr$ is a vector bundle of
2-forms
over $\hat\Ascr$ defined by the condition

$$\hat\Bscr_A\triangleq
\left\{ B\in\Bscr \big|
d_A^*B\in {\rm Im}\left(d_A\big|_{\form{0}}\right)\right\}.\ref\covgauge$$

This choice corresponds to defining
$$\Wscr_A= \left(\H_A\hat\Ascr\ominus \harm{1}\right)
\oplus \hat\Bscr_A.$$

Notice that the orthogonal projection of $\Wscr_A$ on $\H_A \hat\Ascr$
is not any more zero, but it is given by the orthogonal complement of the
harmonic 1-forms.

We refer to the above gauge as the {\it covariant gauge}.

Finally we consider the {\it self-dual gauge.}

We start by considering the projection on self-dual 2-forms \quot
\donaldson{ S. K. Donaldson and  P. B. Kronheimer:
{ \it The Geometry of Four-Manifolds}, Oxford University Press
(Oxford, New York, 1990).}:
$P_+:\form{2}\to {\form{2}}^+.$
We now restrict ourselves
to the open set of connections where the
operator $D_A\triangleq \sqrt{2} P_+ d_A:\form{1}\to\form{2}^+$
is surjective. We denote by $\widetilde{\Ascr}$ such an open
set.

If we define also
$$D_A\triangleq \sqrt{2}d_A:{\form{2}}^+ \to \form{3}$$
and
$$D_A\triangleq d_A:\form{0}\to \form{1},$$
then we can consider the forms that are harmonic with respect  to
the operator $D_A$, for which we use the notation
$\tharm{*}$.

The self-dual
gauge condition is now defined
as follows:
$\eta\in \H_A\widetilde{\Ascr}$ and
$P_+B=0$.

In this way we have defined:
$$\Wscr_A={ {\left(H_A\widetilde{\Ascr}\ominus \tharm{1}\right)
\oplus \Bscr^-}},$$
where $\Bscr^-$ is the space of anti-self-dual elements of $\Bscr$.

Again the orthogonal projection of $\Wscr_A$ on $\H_A \hat\Ascr$
is not  zero, but it is given by the orthogonal complement of the
1-forms that are harmonic with respect to the operator $D_A$.
\blankm
Finally, we consider the equations of motion
relevant to the action functional \bfymeta\ and the corresponding
extrema.

Taking the directional derivatives at $(A,\eta,B)$ in the
directions $(\tau,\mu,C) $ we get the equations
$$\eqalign{-t\langle [\tau,\eta],B-d_A\eta\rangle +\langle
B,*d_A\tau\rangle=0\cr
\langle d_A\mu,B-d_A\eta\rangle=0\cr
t\langle C,B-d_A\eta\rangle+\langle C,*F\rangle=0\cr}$$

Noticing that
$$\langle [\tau,\eta],U\rangle=-\langle\tau,*[*U,\eta]\rangle$$ for a
generic 2-form $U,$ we get
$$\eqalign{t [*B-*d_A\eta,\eta] +d_AB=0
\cr
d_A^*(B-d_A\eta)=0\cr
t (B-d_A\eta)+*F_A =0\cr}$$
Owing  to the Bianchi identity, the second equation is
redundant and the first equation is equivalent to
the Yang--Mills equation $d_A^*F_A=0$. In summary the equations
of motions of \bfymeta\ read
$$\eqalign{d_A^* F_A=0\cr
t (B-d_A\eta)+*F_A =0.\cr}.\ref\motion$$

At $t=1$ the action functional \bfymeta\ reads
$$S_{BFYM,\eta}={1\over 2}
\langle B-d_A\eta+ *F_A,B-d_A\eta +*F_A\rangle-YM(A).\ref
\bfymuno$$
The points $(A,\eta,B)$ satisfying the equation
$$B=d_A\eta-*F_A\ref\min$$
are the minima of  \bfymuno\ once we have fixed
the connection $A$.

When the condition \min\ is satisfied, then \bfymuno\ reduces
to the Yang--Mills action functional (up to a sign).
\sezione{Conclusions}
In this paper we have first discussed the differential geometry of the
tangent bundle $\T P$ of a $G$-principal bundle $P$. It is known
that $\T P$ is a $\T G$-principal bundle.

We have shown that the group of gauge transformations of $\T P$ includes,
as a proper subgroup, the tangent gauge group $\T\Gscr$, namely,
the tangent bundle of the group of gauge transformations on $P$.

The tangent gauge group acts on the tangent bundle of the
space of connections $\Ascr$ of  $P$, and
we have identified this tangent bundle
with a subspace of the space
of connections of $\T P$.

Then we have considered topological and non topological
quantum field theories that admit the tangent gauge group as their symmetry group.

It is  known that:
\blank
\item{A)}
The structure equations and the Bianchi identities for the bundle
$${{P\times \Ascr}\over \Gscr}$$ can be interpreted as the
BRST equations for a topological quantum field theory (see \bs).
\blank
\item{B)}
By taking the restriction along a $\Gscr$-fiber of the bundle $P\times \Ascr$,
one obtains from the above structure equations
the BRST equations for the Yang--Mills theory.
\blank
In this paper we have  first applied the ``tangent
functor" to the above cases A and B. Moreover, a special mixing
of these two cases has been considered by taking
the restriction to the $\Gscr$-orbit of the bundle
$\T P\times \T\Ascr$.

In this way we have
defined a theory that is ``topological" when we consider
the  fields that
are represented
by tangent vectors
to $\Ascr$ and coincides with the Yang--Mills theory when
we consider the connection itself.

The remarkable fact is that, in the above ``semi-topological"
theory, one is able to extend
the BRST operator
to fields represented by 2-forms. In other words one
is allowed to consider field theories that include not only
tangent
vectors to the space of connections, but also tangent vectors to
the space of the curvatures of connections.

The most important non-trivial example
of these field theories is the 4-dimensional
BFYM theory, which, in a separate paper \bfseven,
has been shown to be equivalent to the ordinary
Yang--Mills theory. The fact that 2-forms are among the
fields of this theory can be exploited by considering
vacuum expectation values of observables depending
on iterated loops (loops of loops)
or on imbedded surfaces in a 4-dimensional manifold
\paths.

The gauge fixing problem for the
BFYM theory is discussed in the last section.
In particular the self-dual gauge and the covariant gauge
are compared.

It is rather straightforward to extend all the above calculations to
iterated tangent spaces. Hence we have constructed a BRST complex that
includes---besides the connection $A$---$2^{n+1}-1$
fields which are 1-forms, $2^n$ fields which are 2-forms
and the relevant ghost fields. We are investigating possible
relations between this infinite BRST complex and SUSY models.

Finally, we have shown that (pure) {\it topological}\/ field theories of the $BF$
type do not have a BRST operator related to some Lie algebra cohomology.
In other words the BRST
operator considered in \bs\ does {\it not}\/ extend to 2-forms.
The way of dealing with topological $BF$ theories, from a field-theoretical
point of view, is to resort to the Batalin--Vilkovisky formalism.
\ack{We would like
to thank especially our collaborators of \bfseven---namely,
F. Fucito, M. Martellini, A. Tanzini and M. Zeni---for very useful discussions.
The original motivation for this paper and some of the results
reported here are a result of our common work.
We also thank Jim Stasheff and S. Guerra for reading
the manuscript and
suggesting some corrections.}
\immediate\closeout\fileack
                \par
                \null\blankm
                \centerline{\bf Acknowledgments}
                \blankm
                \input ref.tmp2\vfill\eject
\immediate\closeout\fileref
                \par
                \null\blankm
                \centerline{\bf References}
                \blankm
                \input ref.tmp1\vfill\eject
\bye